\documentclass[aps,prb,twocolumn,superscriptaddress,floatfix,a4]{revtex4}
\usepackage{epsfig,amsmath,amssymb,color}
\begin{document}

\title{Real-time energy dynamics in spin-$1/2$ Heisenberg chains}

\author{Stephan Langer}%
\email{stephan.langer@physik.uni-muenchen.de}%
\affiliation{Department of Physics and Arnold Sommerfeld Center for Theoretical Physics, Ludwig-Maximilians-Universit\"at M\"unchen, D-80333 M\"unchen, Germany}%

\author{Markus Heyl}%

\affiliation{Department of Physics and Arnold Sommerfeld Center for Theoretical Physics, Ludwig-Maximilians-Universit\"at M\"unchen, D-80333 M\"unchen, Germany}%

\author{Ian P. McCulloch}%
\affiliation{School of Physical Sciences, The University of Queensland,
Brisbane, QLD 4072, Australia}

\author{Fabian Heidrich-Meisner}%
\affiliation{Department of Physics and Arnold Sommerfeld Center for Theoretical Physics, Ludwig-Maximilians-Universit\"at M\"unchen, D-80333 M\"unchen, Germany}%

\begin{abstract}
We study the real-time dynamics of the local energy density in the spin-$1/2$ XXZ chain starting from initial states with an inhomogeneous profile of bond energies.
Numerical simulations of the dynamics of the initial states are carried out using the adaptive time-dependent density matrix renormalization group method. 
We analyze the time dependence of the spatial variance associated with the local energy
density to classify the dynamics as
either ballistic or diffusive. Our results are consistent with ballistic behavior both in the massless and the massive phase. We also study the same problem within Luttinger Liquid theory and obtain that energy wave-packets propagate with the sound velocity. We recover this behavior in our
numerical simulations in the limit of very weakly perturbed initial states. \end{abstract}
\maketitle

\section{Introduction}
\label{sec:intro}

\begin{figure}[tb]
  \includegraphics[width=0.6\columnwidth]{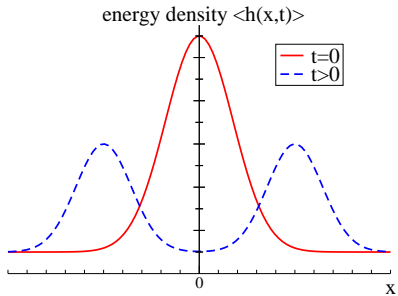}
  \caption{(color online) Sketch of our setup: We prepare initial states with an inhomogeneous distribution of local energies and then study the time evolution of the local energy density.}
 \label{fig:1}
 \end{figure}
 
The understanding of transport properties of low-dimensional systems with strong correlations still poses viable
challenges to theorists. These include, on the one hand, the fundamental problem of calculating transport coefficients
for generic models such as the Heisenberg chain,\cite{zotos-review,hm07a} and on the other hand, the theoretical modeling of experiments 
that typically require the treatment of spin or electronic degrees of freedom coupled to phonons, in particular, in the
case of the thermal conductivity.\cite{boulat07,rozhkov05} Most theoretical work has focussed on the linear-response regime, in which the properties 
of current-current
autocorrelation functions determine transport properties (see Refs.~\onlinecite{zotos-review,hm07a} for a review).

More recently, the out-of-equilibrium properties of one-dimensional systems 
have evolved into an active field of research, one reason being
 recent advances in experiments with ultracold atoms.\cite{bloch08} These have paved the way for studying the dynamics of quantum many-body systems 
that are driven far away from equilibrium in a controlled manner, with little or no coupling to external degrees of freedom.
Much attention has been paid to the question of thermalization, typically studied in so-called quantum quenches (see Ref.~\onlinecite{polkovnikov11}
and references therein).
While global quantum quenches in homogeneous systems usually do not induce any finite net currents (of either spin, energy, or particles),
we will be particularly interested in set-ups that feature finite net-currents.
Such situations are realized in, for instance, the sudden expansion of particles in optical lattices after the removal
of trapping potentials.\cite{schneider11} Further examples are spin and/or particle currents induced by connecting two regions 
with opposite magnetizations or by letting two particle clouds collide (see, for instance, Refs.~\onlinecite{medley11,sommer11a,sommer11b,joseph11}). 

Theoretical work in this context ranges from the expansion dynamics of bosons and fermions in optical lattices\cite{rigol04,rigol05b,kollath05b,hm08a,hm09a,karlsson11,kajala11}
over the dynamics of wave-packets in spin chains,\cite{gobert05,langer09,eisler09,lancaster10a,lancaster10b,mossel10,foster10,santos11,foster11,jesenko11} to the demonstration of 
signatures of spin-charge separation in such set-ups.\cite{kollath05a,polini07} 
In the aforementioned examples, non-equilibrium situations were studied with either finite {\it spin} or {\it particle} currents. 
In our work, we address the {\it energy} dynamics for a model that is 
prototypical for systems with strong correlations, namely the spin-1/2 XXZ chain: 
\begin{align}
H_{XXZ}&=\sum_{i=1}^{L-1}h_i\notag\\&:= J\sum_{i=1}^{L-1} \lbrack\frac{1}{2}(S_{i}^+S_{i+1}^{-}+H.c.) + \Delta S^z_iS^z_{i+1} \rbrack\, , \label{eq:xxz}
\end{align}
where $S_i^{\mu}$ and $\mu=x,y,z$ are the components of a spin-$1/2$ operator acting on site $i$
and $S^{\pm}_i$ are the corresponding lowering/raising operators. The global energy scale is set by the exchange coupling $J$, $\Delta$ is the exchange anisotropy in the $z$-direction, and $L$ denotes the number of sites.  
Equation~\eqref{eq:xxz} describes either interacting quantum spins or, via the Jordan-Wigner transformation,\cite{jowi} spinless fermions.

Specifically, we follow the time-evolution of the local energy density $\langle h_i\rangle $ starting from initial states that are far from the ground state of Eq.~\eqref{eq:xxz} and that feature an inhomogeneous profile in the local energy density (see Fig.~\ref{fig:1} for a sketch).
We emphasize that, in the main part of our work, we
choose the initial conditions such that {\it only} finite energy currents exist, whereas the spin (particle) density is constant during the time evolution, hence all spin (particle) currents vanish.
Obviously, an initial state with an inhomogeneous spin density profile leads to both finite spin {\it and} energy currents, and we revisit this case, previously studied in Refs.~\onlinecite{langer09} and \onlinecite{jesenko11}, as well.

Our work is motivated by and closely related to a specific experiment on a spin-ladder material. 
Many low-dimensional quantum magnets are known to be very good thermal conductors with heat 
predominantly carried by magnetic excitations at elevated temperatures.\cite{hess07,sologubenko07a} Examples for materials that exhibit particularly large thermal conductivities
are (Sr,La,Ca)$_{14}$Cu$_{24}$O$_{41}$ (Refs.~\onlinecite{hess01} and \onlinecite{sologubenko00}) and SrCuO$_2$ (Ref.~\onlinecite{hlubek10}). While these experiments are carried
out under  steady-state conditions and in the regime of small external perturbations, more recently, time-resolved measurements have been performed
on La$_{9}$Ca$_{5}$Cu$_{24}$O$_{41}$ (Ref.~\onlinecite{otter09}). For this spin ladder material, two approaches have been implemented: A time-of-flight 
measurement, in which one side of the sample is heated up with a laser pulse and the time-dependent response is recorded on the other side. Second,
a non-equilibrium local heat distribution was generated in the surface of the material by shining laser light on it. It is possible to record the heat dynamics 
via thermal imaging that uses the response of an excited thin fluorescent layer placed on top of the spin ladder material. 

It is the latter case that we mimic in our work: The time evolution of local energy densities induced by inhomogeneous initial distributions. We utilize
the time-dependent density matrix renormalization group (tDMRG)\cite{daley04,white04,vidal04,schollwoeck05,schollwoeck11} technique. 
It allows us to simulate the dynamics of pure states whereas in the experiment, temperature likely plays a role. 
Our work thus addresses qualitative aspects in the first place, while a direct comparison with experimental results is beyond the scope of this study: 
The goal is to demonstrate that in a spin-1/2 chain described by Eq.~\eqref{eq:xxz}, the energy dynamics is ballistic, 
irrespective of how far from equilibrium the system is and also irrespective of the presence or absence of excitation gaps. 
To this end, we use the same approach as in Ref.~\onlinecite{langer09}:
We classify the dynamics based on the behavior of the spatial variance $\sigma_E^2(t)$ of the local energy density:
The ballistic case is $\sigma^2_E(t)\sim t^2$, whereas diffusion implies $\sigma^2_E\sim t$.
Our main result for the XXZ chain, based on numerical tDMRG simulations, is that energy propagates ballistically at sufficiently long times, independently
of model parameters (such as $\Delta$). One can then interpret the prefactor $V_E$ in
$\sigma^2_E(t)=V_E^2 t^2$ as a measure of the average velocity of excitations contributing to the expansion.
The velocity $V_E$ can be calculated analytically and exactly in non-interacting models, which (in the absence of impurities or disorder) typically have ballistic dynamics, and we consider two examples: (i) the noninteracting
limit of the XXZ-Hamiltonian ($\Delta=0$), i.e., spinless fermions and (ii) the Luttinger liquid, which is the
universal low-energy theory in the continuum limit of Eq.~\eqref{eq:xxz} for $|\Delta|<1$. We show that our tDMRG results
agree with the exactly known expansion velocity $V_E$ in these two examples.

Our main result, namely the numerical observation of $\sigma_E^2(t)\sim t^2$ independently of initial conditions or model parameters such as the exchange anisotropy $\Delta$,
is consistent with the qualitative picture derived from linear-response theory. Within that theory transport properties of the XXZ chain have intensely been studied in recent years, both the energy\cite{zotos97,kluemper02,sakai03,hm02,hm03} and the spin transport.
\cite{narozhny98,zotos99,hm03,zotos04,prelovsek04,benz05,jung06,hm07a,sirker09,steinigeweg09,grossjohann10,sirker11,steinigeweg11,steinigeweg11a,zotos-review,prosen11,znidaric11,herbrych11} 
\cite{,}
Ballistic dynamics is associated with the existence of non-zero Drude weights. Since the total energy
current of the anisotropic spin-1/2 chain is a conserved quantity for all $\Delta$, the thermal conductivity $\kappa(\omega)$ diverges in the zero-frequency limit
and is given by $\mbox{Re}\, \kappa(\omega)=D_{E}\delta(\omega)$, where $D_E$ is the thermal Drude weight.\cite{zotos97,kluemper02,sakai03,hm02}
This behavior is different from the spin conductivity $\sigma({\omega})$. This quantity takes the form $\mbox{Re}\, \sigma(\omega)=D_{s}\delta(\omega)$
only at the noninteracting point $\Delta=0$, whereas for $0<\Delta\leq 1$, many numerical studies\cite{zotos-review,hm07a,prosen11,znidaric11} indicate $D_s(T>0)>0$, with a 
\emph{finite weight at finite frequencies}, though. Therefore, for $0<\Delta\leq 1$, $\mbox{Re}\,\sigma(\omega)=D_s\delta(\omega)+\sigma_{\mathrm{reg}}(\omega)$. Recent field-theoretical and numerical work suggests that the regular part $\sigma_{\mathrm{reg}}(\omega)$ of $\sigma(\omega)$ 
in massless phases is consistent with diffusive behavior.\cite{sirker09,grossjohann10,sirker11} 
A finite value of the current-current correlation function in the long time limit is associated with a finite Drude weight. 
Finite Drude weights can be traced back to the existence of conservation laws,\cite{zotos97,prosen11}
and in consequence, a potential relation between integrability\cite{caux11} and ballistic behavior - in the sense of non-zero
Drude weights - has been intensely discussed (see, e.g., Refs.~\onlinecite{zotos-review,hm07a,hm03,sirker11,prosen11,znidaric11} and
further references cited therein).
Very recently, Prosen has presented results that provide a lower bound to the spin Drude weight that is non-zero for $\Delta<1$.\cite{prosen11} This is in qualitative agreement with earlier exact diagonalization studies.\cite{narozhny98,hm03,hm07a}
The particular point $\Delta=1$ is still discussed controversially:\cite{zotos99,hm03,benz05,sirker09,prosen11,znidaric11,herbrych11} First, no finite lower bound to the Drude weight is known,\cite{prosen11} and second, the qualitative results
of exact diagonalization studies seem to depend on details of the extrapolation of finite-size data to the thermodynamic limit and the statistical ensemble that is considered.\cite{narozhny98,hm03,herbrych11} 

Our approach that analyzes the time-dependence of spatial variances, albeit restricted to the analysis of densities, is numerically easily tractable
and is an alternative to the numerically cumbersome evaluation of current correlation functions.
tDMRG has, for instance, been applied to evaluate current-current autocorrelation functions in the thermodynamic limit.\cite{sirker09} However, the accessible time scales are quite limited $(t\sim10/J$), making an unambiguous interpretation of the results difficult and the approach is not applicable to non equilibrium.
 Our approach allows us, at least in principle, to
study the entire regime of weakly perturbed states to maximally excited ones.
 An earlier analysis of spin-density wave packets in various spin models has yielded the following picture (all based on the time-dependence of the spatial variance):\cite{langer09}
In massless phases, ballistic dynamics is seen, whereas in massive ones, examples of diffusive dynamics have been identified. It is important to stress
that the observation of a variance that increases linear in time is a \emph{necessary} condition 
for the validity of the diffusion equation. 

Finally, to complete the survey of related literature, recent studies have addressed steady-state spin and energy transport in open systems
coupled to baths, with
no restriction to the linear-response regime.\cite{prosen09,benenti09,benenti09a,znidaric11,jesenko11} 
These studies suggest spin transport to be ballistic in the gapless phase of the XXZ spin chain and to be diffusive in the gapped phase 
with a negative differential conductance at large driving strengths. 
The heat current has been addressed in Ref.~\onlinecite{prosen09} where Fourier's law has been validated for the Ising model in a tilted field. 

A by-product of any tDMRG simulation is information on the time-evolution
of the entanglement entropy. While this is not directly related to this article's chief case, it nevertheless provides valuable information on
the numerical costs of tDMRG simulations. Qualitatively, speaking (see the discussion in Ref. \onlinecite{schollwoeck11} and references therein), the faster the
entanglement growth is, the shorter are the time scales that can be reached
with tDMRG. We here show that the quenches studied in this work generate a mild
logarithmic increase of entanglement, which is why this problem is very well
suited for tDMRG. Such a behavior is typical for so-called local quenches.\cite{calabrese07} This result might be useful for tDMRG practitioners.

 \begin{figure*}[tb]
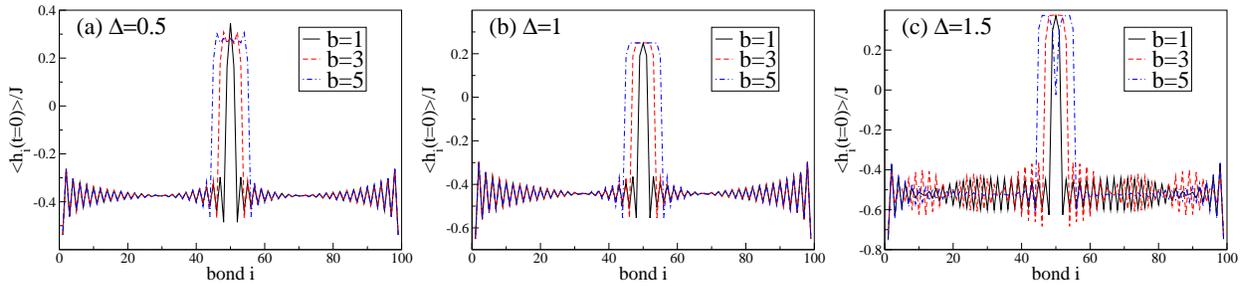

  \includegraphics[width=0.3\textwidth]{fig2a.eps}
  \includegraphics[width=0.3\textwidth]{fig2b.eps}
   \includegraphics[width=0.3\textwidth]{fig2c.eps}
  \caption{(color online) Profile of the local energy density $\langle h_i\rangle$ in the initial states induced by a $J_i$ quench for $b=1,3,5$ [compare Eq.~\eqref{eq:block}] 
for (a) $\Delta=0.5$, (b) $\Delta=1$ and (c) $\Delta=1.5$. In all cases, the system forms a region with ferromagnetic nearest-neighbor spin correlations in the middle of the chain. 
In the regions with antiferromagnetic $J_i>0$, the local energy density oscillates, reflecting the antiferromagnetic nearest-neighbor correlations.} 
\label{fig:xxzinit}
 \end{figure*} 
 
This paper is organized as follows: First, we introduce the model and the quantities used in our analysis in Sec.~\ref{sec:model}. Section~\ref{sec:boson} 
reviews the framework of bosonization, which is applied in Sec.~\ref{sec:bosdyn} to give an analytical derivation of ballistic spin and energy dynamics in the 
low-energy case, valid in the massless phase of Eq.~\eqref{eq:xxz}.
Sections~\ref{sec:4} and \ref{sec:5} contain our numerical results. 
First, we study the energy dynamics in the absence of spin currents in Sec~\ref{sec:4}. To this end, we generate an initial state consisting of a variable number of ferromagnetic bonds in the center of an antiferromagnetic chain. We calculate the time evolution of these states under Eq.~\eqref{eq:xxz} finding ballistic energy dynamics independent of the phase and the strength of the perturbation. To supplement these findings we derive an observable, which depends on the local currents, and whose expectation value is time-independent whenever $\sigma_E^2(t)\sim t^2$. The numerical calculation of this quantity indicates ballistic dynamics as well. Section~\ref{sec:5} revisits the scenario of Ref.~\onlinecite{langer09} where local spin and energy currents are present during the dynamics as we start from states with an inhomogeneous spin density. In that case the energy density shows ballistic dynamics in the massless phase with a velocity matching the bosonization result in the limit of small perturbations. In the massive phase we observe a different behavior of the two transport channels, i.e., ballistic energy dynamics while the spin dynamics looks diffusive.\cite{langer09} 
Finally, we summarize our findings in Sec.~\ref{sec:sum}. Additionally, we discuss the entanglement growth induced by coupling two regions with an opposite sign 
of the exchange coupling in the Appendix.


\section{Setup and definitions}
\label{sec:model}

\subsection{Preparation of initial states and definition of spatial variance}

In this work, we focus on spin-$1/2$ XXZ chains of a finite length $L$ given by Eq.~\eqref{eq:xxz} where our goal is to study the dynamics of an inhomogeneous distribution of the local energy density originating from a local quench of system parameters. The inhomogeneous distributions are 
generated by preparing the system in the respective ground states of the following Hamiltonians that are perturbations of $H_{XXZ}$ from Eq.~\eqref{eq:xxz}.
First, 
\begin{equation}
	H_{\mathrm{init}}^J=\sum_{i=1}^{L-1} \frac{J_i}{J} h_i\,,\label{eq:hp}
\end{equation}
where $h_i$ is defined in Eq.~\eqref{eq:xxz}, and second, 
\begin{equation}
H^B_{\mathrm{init}}=H_{\mathrm{XXZ}}-\sum_iB_iS^z_i \,,\label{eq:gaussB0}
\end{equation}
where
\begin{equation}
B_i=B_0\,e^{\frac{-(i-L/2)^2}{2\sigma_0^2}}\,.
\end{equation}
In the first case, we quench site-dependent exchange couplings. In this scenario we obtain initial states with large local energy densities. Typical initial states that are ground 
states of Eq.~\eqref{eq:hp} are shown in Fig.~\ref{fig:xxzinit}: These states have $b$ bonds with ferromagnetic $J_i<0$ in the center
while the rest has antiferromagnetic $J_i>0$. We refer to this setup as the $J_i$ quench.

In the second case, the dynamics is driven by an inhomogeneous spin density, enforced by an external magnetic field applied in the initial state. This allows us to generate smooth spatial perturbations of $\langle h_i\rangle$ with small differences in energy compared to the ground state of Eq.~\eqref{eq:xxz}. We refer to this setup as the $B_0$ quench. A more detailed discussion of the initial states generated by a $J_i$ quench will be given in Sec.~\ref{sec:4a}. The $B_0$ quench was introduced in detail in Ref.~\onlinecite{langer09}.

The definition of the local energy density from the Hamiltonian Eq.~(\ref{eq:xxz}) is not unambiguous. For instance, it is always possible to add local terms to the
 Hamiltonian whose total contribution by summation over all lattice sites vanishes. 
 However, this seeming ambiguity can be resolved up to constants by requiring that any block of adjacent 
 lattice sites $\sum_{i=l}^m h_i$ is Hermitian and yet to have the same structure as the total Hamiltonian $H$. 
 These details seem to be rather specific, yet for the definition of the appropriate local energy density within the Luttinger 
 liquid description, see below, these formal considerations are important. For the XXZ chain the local energy density is 
 therefore determined by the bond energies $\langle h_i\rangle$. 

To classify the dynamics of a density $e_i$ we study its spatial variance. 
\begin{equation} \sigma_E^2(t)=\sum_{i=1}^{L-1} (i-\mu)^2 e_i(t), \label{eq:sigma} \end{equation} 
where $\mu$ is the first moment of $e_i$. The $e_i$ are the normalized distribution linked to the energy density via
\begin{equation}
	e_i=\delta E^{-1} \langle \tilde{h}_i \rangle \label{eq:dist}
\end{equation}
where $\langle \tilde{h}_i \rangle=\langle h_i\rangle-\langle h_i\rangle_0$ denotes the expectation value of $h_i$ in the initial state shifted by the ground state expectation value $\langle h_i\rangle_0=\langle\psi_{0}| h_i|\psi_{0}\rangle$. 
\begin{equation}
\delta E:=E_{\mathrm{init}}-E_0=\sum_i\langle \tilde{h}_i\rangle
\end{equation}
is the energy difference between the initial state $| \psi_{\mathrm{init}}\rangle$ [i.e., the ground state of either $H^J_{\mathrm{init}}$ or  $H^B_{\mathrm{init}}$ ] and the ground state $| \psi_0\rangle$ 
of Eq.~\eqref{eq:xxz}, both energies measured with respect to the {\it unperturbed} Hamiltonian from Eq.~\eqref{eq:xxz}:
\begin{equation}
E_0=\langle \psi_0 | H_{\mathrm{XXZ}}| \psi_0\rangle; \enspace E_{\mathrm{init}}=\langle \psi_{\mathrm{init}} | H_{\mathrm{XXZ}}| \psi_{\mathrm{init}}\rangle\,.
\end{equation}

On physical grounds, the energy density should be normalized by the
amount of energy transported by the propagating perturbation. This
is well approximated by the energy difference $\delta E$ between the
initial state and the ground state of Eq.~\eqref{eq:xxz}, as we have verified in many examples.
In some cases, though, the propagating energy is, on a quantitatively
level, better described by
estimating the area under the perturbations, as $\delta E$ may also
contain contributions from static deviations from the ground state
bond-energies in the background. Nevertheless, $\delta E$
does not depend on the overall zero of energy and is an obvious measure
of how far the system is driven away from the ground state. This, all together,
justifies our definition of the $e_i$.

To remove static contributions depending only on the initial distribution $e_i(t=0)$, we subtract $\sigma_E^2(t=0)$ and study $\delta \sigma_E^2(t):=\sigma_E^2(t)-\sigma_E^2(0)$. $\delta\sigma_E^2(t)\sim (V_E t)^2$ is expected to grow quadratically in time in the case of ballistic behavior, where $V_E$ has the dimensions of a velocity. For diffusive behavior, we expect, from the fundamental solution of the diffusion equation,\cite{chandrasekhar43} that $\delta \sigma_E^2(t)\sim D t$ grows linearly in time, where $D$ is the diffusion constant (see, e.g, the discussion in Ref.~\onlinecite{langer09}). 
Within linear response 
theory the diffusion constant can be related to transport coefficients via Einstein 
relations, see e.g. Ref. \onlinecite{steinigeweg10}. 
To be clear the observation of $\delta\sigma_E^2\sim t^2$ or $\delta\sigma^2_E\sim t$ is a \emph{necessary} conditions for the respective type of dynamics and time-dependent crossovers are possible. 

\subsection{Spatial variance in the non-interacting case}
For pedagogical reasons and to guide the ensuing discussion we next calculate the 
spatial variance in the non-interacting limit of Eq.~\eqref{eq:xxz}, i.e., at $\Delta=0$.
Using the Jordan-Wigner transformation, we can write the Hamiltonian as
\begin{equation}
H=\frac{J}{2} \sum_i (S^+_i S_{i+1}^-+h.c.)= -\frac{J}{2} \sum_i (c_{i}^{\dagger}c_{i+1} +h.c.)\,, 
\end{equation}
where $c^{\dag}_i$ creates a spinless fermion on site $i$.
A subsequent Fourier transformation diagonalizes the Hamiltonian:
\begin{equation}
H=\sum_k \epsilon_k c_k^{\dagger}c_k\,.
\end{equation}
Since we will compare with numerical results on systems with open boundary conditions, 
we obtain
\begin{equation}
\epsilon_k = -J \cos(k); \quad k=\frac{\pi n}{L+1}; \quad n=1,\dots, L\,.
\end{equation}

Next, we compute
$$
\delta \sigma_E^2(t) = \sum_i e_i(t) (i-i_0)^2 - \sum_i e_i(t=0) (i-i_0)^2
$$
with $e_i$ from Eq.~\eqref{eq:dist} and $h_i=-J(c_{i}^{\dagger}c_{i+1} +h.c.)/2$.
By expressing $c_i^{(\dagger)}$ through their Fourier transform and by plugging in the time evolution
of $c_{k}^{(\dagger)}$, we finally obtain, after straightforward calculations:
\begin{equation}
\delta \sigma_E^2(t) = V_E^2\, t^2\,,\label{eq:sE}
\end{equation}
i.e., ballistic dynamics independently of the initial state. Terms linear in $t$ will be absent  
if in the initial state, the density is  symmetric with respect to its first moment, i.e., $e_{\mu+\delta}=  e_{\mu-\delta}$ 
and if the wave packet has no  finite center-of-mass momentum at $t=0$ already. In the remainder of the paper we will work under these two additional assumptions  
that are valid for all initial states considered in our work. 
The prefactor $V^2_E$ is given by:
\begin{equation}
V^2_E = \frac{1}{\delta E} \sum_k \epsilon_k v_k^2 \delta n_k \,,\label{eq:v2}
\end{equation}
where $v_k=\partial \epsilon_k/ \partial k$ and 
$$\delta n_k =n_k^{\mathrm{init}}-n_k $$ 
is the 
difference between the momentum distribution function (MDF) in the initial state
and the one in the ground state of Eq~\eqref{eq:xxz}. Since we use open boundary conditions, we compute $n_k$ from
\begin{equation}
 n_k=\langle c^{\dagger}_k c_k\rangle:=\frac{2}{L+1}\sum_{r,r^{\prime}} \sin{(kr)}\sin{(kr^{\prime})} \langle c^{\dagger}_r c_{r^{\prime}} \rangle\,. \label{eq:mdf}
\end{equation}
We can also express $\delta E$ via $\delta n_k$:
$$
\delta E= \sum_k \epsilon_k\, \delta n_k\,.
$$
The expression Eq.~\eqref{eq:v2} suggests that $V_E$ is the average velocity of excitations contributing to the propagation of the wave packet.
Characteristic for ballistic dynamics, $V^2_E$ is fully determined by the initial conditions through 
$\delta n_k$.

For completeness, we mention that an analogous calculation can be done for the
spatial variance $\sigma_S$ of the spin density. 
This quantity is defined as
\begin{equation}
\sigma_S^2(t):= \frac{1}{\mathcal{N}}\sum_{i=1}^{L}(i-\mu)^2\cdot \langle S^z_i(t)+1/2\rangle\,.\label{eq:sigmaS}
\end{equation}
The normalization constant $\mathcal{N}$ measures the number of propagating particles.
The spin density is, in terms of spinless fermions:
$$
S^z_i = c^{\dagger}_i c_{i}-1/2=n_i-1/2\,.
$$
The result for the spatial variance of the spin density is 
\begin{equation}
\delta \sigma_S^2(t)=\sigma_S^2(t)-\sigma_S^2(0)=V_S^2 t^2\, \label{eq:sigmas}
\end{equation}
with 
\begin{equation}
V^2_S = \frac{1}{\mathcal{N}} \sum_k v_k^2 \delta n_k \,.\label{eq:vs}
\end{equation}

Although we started from the Hamiltonian for $\Delta=0$, we stress that Eqs.~\eqref{eq:sE}, \eqref{eq:sigmas}, \eqref{eq:v2}, and \eqref{eq:vs}
are valid for any dispersion relation $\epsilon_k$, irrespective of the presence of a gap, provided that $k$ has the meaning of a momentum.


\subsection{Energy current}
\label{sec:currents}
Another aspect worth noting is that the time-evolving state carries a nonzero energy current, a situation that 
usually does not appear in the case of a global quench. 
From the equation of continuity for the energy density one can derive the well-known expression for the local energy current operator\cite{zotos97}
\begin{equation}
j_i^E=J^2\tilde{\vec{S}}_{i-1}\cdot(\vec{S_i}\times\tilde{\vec{S}}_{i+1})\,,\label{eq:jenergy}
\end{equation}
 where $\tilde{\vec{S}}=(S^x,S^y,\Delta S^z)$. 
With periodic boundary conditions, the total current $J_E=\sum_i j_i^E$ is a conserved quantity, i.e., $\lbrack H, J_E\rbrack =0 $ (see Ref.~\onlinecite{zotos97}).
On a system with open boundary conditions such as the ones that are well-suited for DMRG, this property is lost, yet the dynamical conductivity
still has a quasi-Drude peak at very low frequencies, reminiscent of the true Drude peak $Re\kappa(\omega) =D_{E} \delta(\omega)$ of a system with periodic boundary
conditions.\cite{rigol08} The latter form is recovered on a system with open boundary conditions as $L\rightarrow \infty$ (Ref.~\onlinecite{rigol08}), showing
that ballistic dynamics due to the existence of globally conserved currents can still be probed on systems with open boundary conditions.

To connect the local energy currents to the spatial variance of the time-dependent density one can rewrite the time derivative of $\sigma_E^2(t)$ using the equation of continuity, assuming no current flow to sites at the boundary (this assumption is justified in our examples as long as we restrict ourselves to times before reflections occur at the boundary in our simulations):
\begin{align}
\partial_t\sigma^2_E(t)&\sim\sum_{r=1}^L (r-\mu)^2\partial_t \langle h_r(t)\rangle \notag\\&=-\langle j^E_1\rangle+\sum_{r=1}^L\, (2r-2\mu+1)\langle j^E_r(t)\rangle\,.
\end{align}
 If $\sigma_E^2(t)=V_E^2 t^2+b$ and $\mu\neq\mu(t)$, then using $\langle J_E\rangle=0 $ leads to:
 \begin{equation}
\sum_{r=1}^L \,r\,\partial_t\langle j^E_r(t)\rangle \sim\frac{1}{2}\partial^2_t\sigma_E^2(t)=V_E^2=\mbox{const}\,.
 \label{eq:Jstar0}
 \end{equation}
If we interpret this equation as an operator equation, then we see that
we can define a quantity $J_E^*$ via:
 \begin{equation}
J_E^*=\sum_{r=1}^L \,r\,\partial_t j^E_r\,. 
 \label{eq:Jstar}
 \end{equation}
If for a given initial state and over a certain time window, $\langle J^*_E(t)\rangle=$~const, then we have identified
a regime with ballistic dynamics, $\delta \sigma_E^2(t)\sim t^2$. If $\langle J^*_E(t)\rangle=$const holds for all times
and initial states, then $J^*_E$ is a conserved quantity, $\lbrack H,J_E^*\rbrack=0$. This is the case at $\Delta=0$, the non-interacting limit of Eq.~\eqref{eq:xxz}, where $\langle J_E^*\rangle=V_E^2\,\delta E$ from Eq.~\eqref{eq:v2}.

We emphasize that we have here identified a operator that connects the phenomenological observation of a quadratic increase of $\sigma^2_E(t)$ 
to the local energy currents. In ballistic regimes, its expectation value becomes stationary.

For completeness, we mention an analogous result in the diffusive regime where $\sigma_E^2\sim t$. Then, expectation values of
the operator
\begin{equation}
J_E^D =\sum_{r=1}^L (r-\mu) j_r^E (t)
\end{equation} 
are time independent. Obviously, similar expressions can be written down for the spatial variance associated with the 
spin density.


\section{Propagating energy and spin wave-packets in a Luttinger liquid}
\label{sec:ll}

In the gapless phase, i.e., for $|\Delta| < 1$, the low-energy and low-momentum properties of the XXZ chain can be described by an effective Luttinger liquid theory.~\cite{solyom79} In the following we want to analyze the energy density and the spin dynamics of the XXZ chain in this exactly solvable hydrodynamic limit. Specifically, we show that at least asymptotically for large times, the spatial variance always grows quadratically both in the case of spin and energy dynamics. In addition, we work out the precise dependence of the prefactor in front of the $t^2$ increase of the spatial variance on system parameters. Since our DMRG results to be presented in Secs.~\ref{sec:4} and \ref{sec:5} show that $\sigma_E^2(t)\sim t^2$ at any $\Delta$, we did not investigate the influence of marginally relevant perturbations at $\Delta=1$ on the wave-packet dynamics. In passing, we mention
that in the massive phase, where the appropriate low-energy theory is the sine-Gordon model, the expansion velocity could also be derived at the Luther-Emery point 
(this case was studied, in, e.g. Refs.~\onlinecite{foster10,foster11}).

\subsection{Bosonization of the anisotropic spin-1/2 chain}
\label{sec:boson}

The Hamiltonian Eq.~(\ref{eq:xxz}) can be mapped onto a system of interacting spinless fermions via the Jordan-Wigner transformation.\cite{jowi} Within a hydrodynamic description in terms of a linearized fermionic dispersion relation the Hamiltonian can be represented in terms of a Luttinger liquid theory (LL)
\begin{eqnarray}
	H_{LL} = \frac{u}{4} \int \frac{dx}{2\pi} \: \left[ K \left[ \rho_L-\rho_R\right]^2 + \frac{1}{K} \left[ \rho_L + \rho_R \right]^2 \right] \label{eq:LL_Hamiltonian} 
\end{eqnarray}
using the notation of Ref.~\onlinecite{vondelft98}. The sum of the two left- and right-mover densities $\rho_{L}(x)+\rho_R(x)$ of the spinless Jordan-Wigner fermions is proportional to the continuum approximation of the local magnetization $S_i^z$ up to a constant. The sound velocity $u$ can be related to the parameters of the XXZ chain in Eq.~(\ref{eq:xxz}) via the group velocity\cite{cloizeaux66}
\begin{equation}
u=v_g=J \frac{\pi}{2}\frac{ \sin(\nu)}{\nu}\,, \label{eq:vg}
\end{equation}
 with $ \cos \nu=\Delta$. Similarly, the Luttinger parameter $K$ is given by the relation $K=\pi/[2(1-\nu)]$. In the noninteracting case $\Delta=0$ we have $K=1$ and $u=J$.

\subsection{Ballistic dynamics in the gapless phase} 
\label{sec:bosdyn}
Within the Luttinger liquid description for $\Delta<1$, an initially inhomogeneous local energy density profile 
always propagates ballistically independently of the details of the perturbation as can be seen from general arguments. For the effective low-energy Hamiltonian the probability distribution $e(x,t)$ associated with the local energy density is given by 
\begin{equation}
e(x,t)=\mathcal{E}^{-1}\langle \psi_{\mathrm{init}} | \hat{h}(x,t) | \psi_{\mathrm{init}} \rangle\,, 
\end{equation}
where $|\psi_{\mathrm{init}}\rangle$ is the initial state, 
\begin{align}
\hat h(x)=u(K+K^{-1})/(8\pi) \sum_\eta \partial _x \varphi_\eta^\dag(x) \partial_x \varphi_\eta(x)\notag\\ -u(K-K^{-1})/(8\pi) (\partial_x \varphi_L^\dag(x) \partial_x \varphi_R^\dag(x)\\+\partial_x \varphi_R(x) \partial_x \varphi_L(x)) \notag
\end{align}
 and 
\begin{equation}
 \mathcal{E}=\int dx \: \langle \psi_{\mathrm{init}} | \hat h(x,t=0) |\psi_{\mathrm{init}} \rangle\,.
\end{equation} 
 For the exact definition of the fields $\varphi_\eta^{(\dag)}$, see, e.g., Ref.~\onlinecite{vondelft98}. The local energy density operator consists of decoupled left- and right-moving contributions in the basis in which the Hamiltonian for the time evolution is diagonal. This allows for a separation of $e(x,t)$ into left- and right-moving contributions which both propagate with the sound velocity $v_g$: $e(x,t)=e_L(x+v_g t,t=0)+e_R(x-v_gt,t=0)$. 
 
 Assuming a $L \leftrightarrow R$ symmetry in the initial state, i.e., a state with zero total momentum, one obtains for the variance from Eq.~\eqref{eq:sigma}
\begin{equation}
	\delta \sigma_E^2(t)=\sigma_E^2(t)-\sigma_E^2(t=0)=(V_Et)^2\
\end{equation}
for all times $t$ with $V_E=v_g=u$. This results can also be obtained from evaluating 
Eq.~\eqref{eq:v2} in the continuum limit. 

In the case of an initial $L\leftrightarrow R$ asymmetry in the initial state we get $\delta \sigma_E^2(t)\rightarrow (v_gt)^2$ for $t\to \infty$, but the short time behavior may differ. Thus, within the validity of a Luttinger liquid description the energy transport is always ballistic for all initial conditions. This is evident from a physical point of view as all excitations propagate with exactly the same velocity $v_g$, the left-movers to the left and the right-movers to the right. Note that the applicability of a Luttinger liquid description is manifestly restricted to cases in which the initial energy density profile is a smooth one 
in the sense that the associated excitations do not feel the nonlinearity of the fermionic dispersion relation. 
Thus, the time-evolution starting from initial profiles such as the ones shown in Fig.~\ref{fig:xxzinit} are beyond the scope of this low-energy theory.

In analogy to the above arguments, the dynamics of spin-density wave-packets is also ballistic in the $XXZ$ chain for $\Delta<1$ in the Luttinger liquid limit. In the bosonic theory, the spin density is proportional to $\rho_L(x)+\rho_R(x)$ up to a constant, see Sec.~\ref{sec:boson}. The associated probability distribution $\rho(x,t)= \mathcal{Q}^{-1} \langle \rho_L+\rho_R \rangle/2\pi $, with $\mathcal{Q}=\int dx \langle \rho_L+\rho_R\rangle/2\pi$, can again be separated into a left- and a right-moving contribution, i.e., 
\begin{equation}
\rho(x,t)=\rho_L(x+v_gt,t=0)+\rho_R(x-v_gt,t=0)\,.
\end{equation}
Thus, similar to the case of the energy dynamics, one finds ballistic behavior for $|\Delta|<1$ consistent with the numerical results of Ref.~\onlinecite{langer09}.

\begin{figure}[tb]
  \includegraphics[width=0.7\columnwidth]{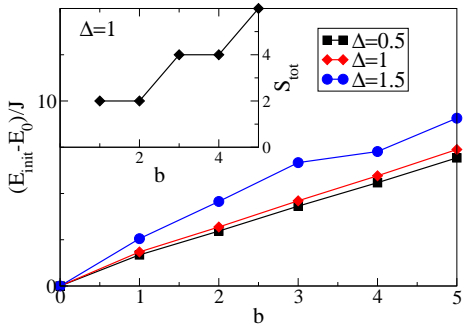}
  \caption{(color online) Energy difference $\delta E$ between the initial state and the ground state for the $J_i$ quench as a function of $b$ for $\Delta=0.5,1,1.5$. The inset shows the hierarchy of states with increasing total spin which appear as initial states when the total spin is a good quantum number, i.e., at $\Delta=1$.}
 \label{fig:energies}
 \end{figure}
 
\begin{figure*}[tb]
  \includegraphics[width=0.3\textwidth]{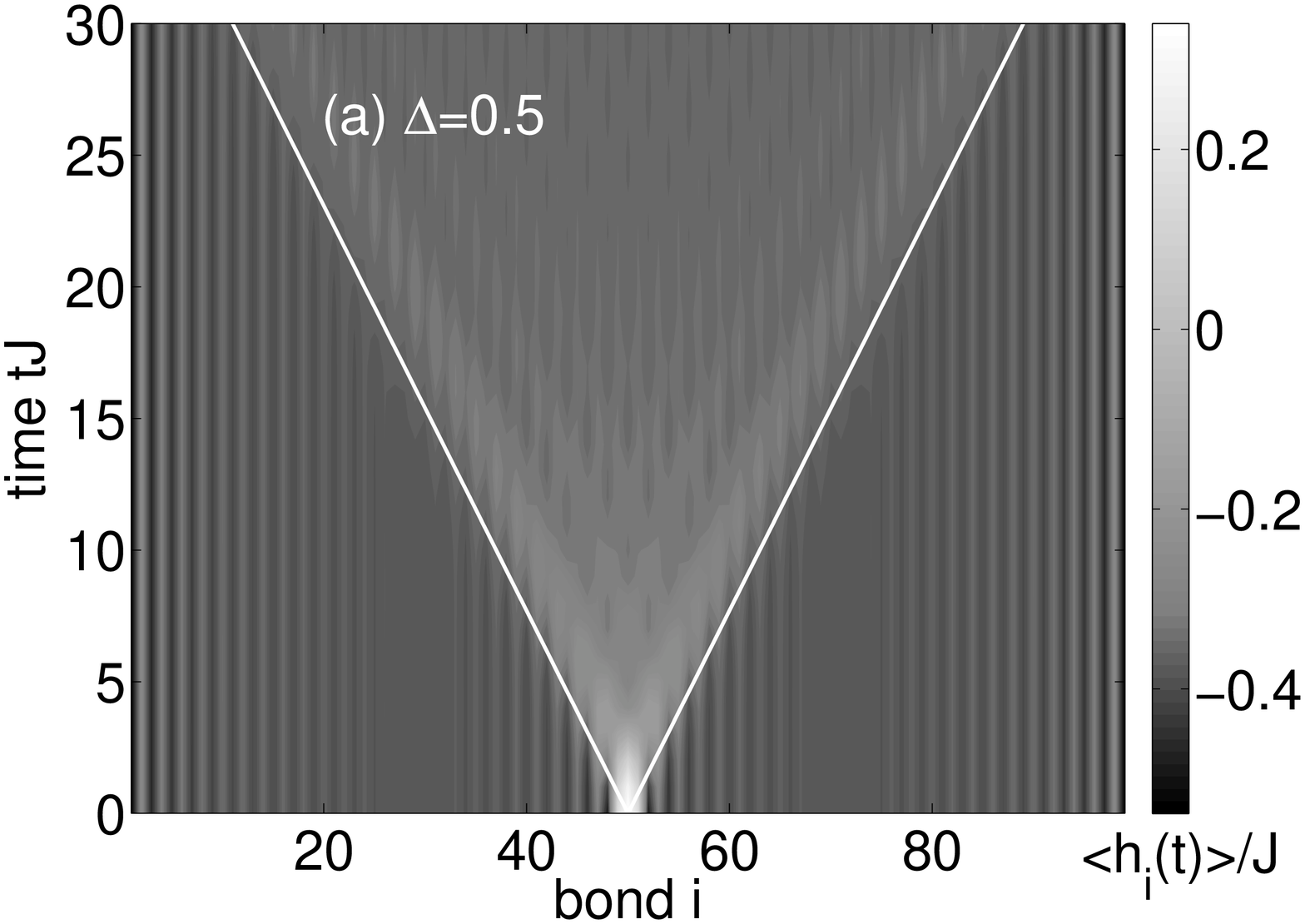}
  \includegraphics[width=0.3\textwidth]{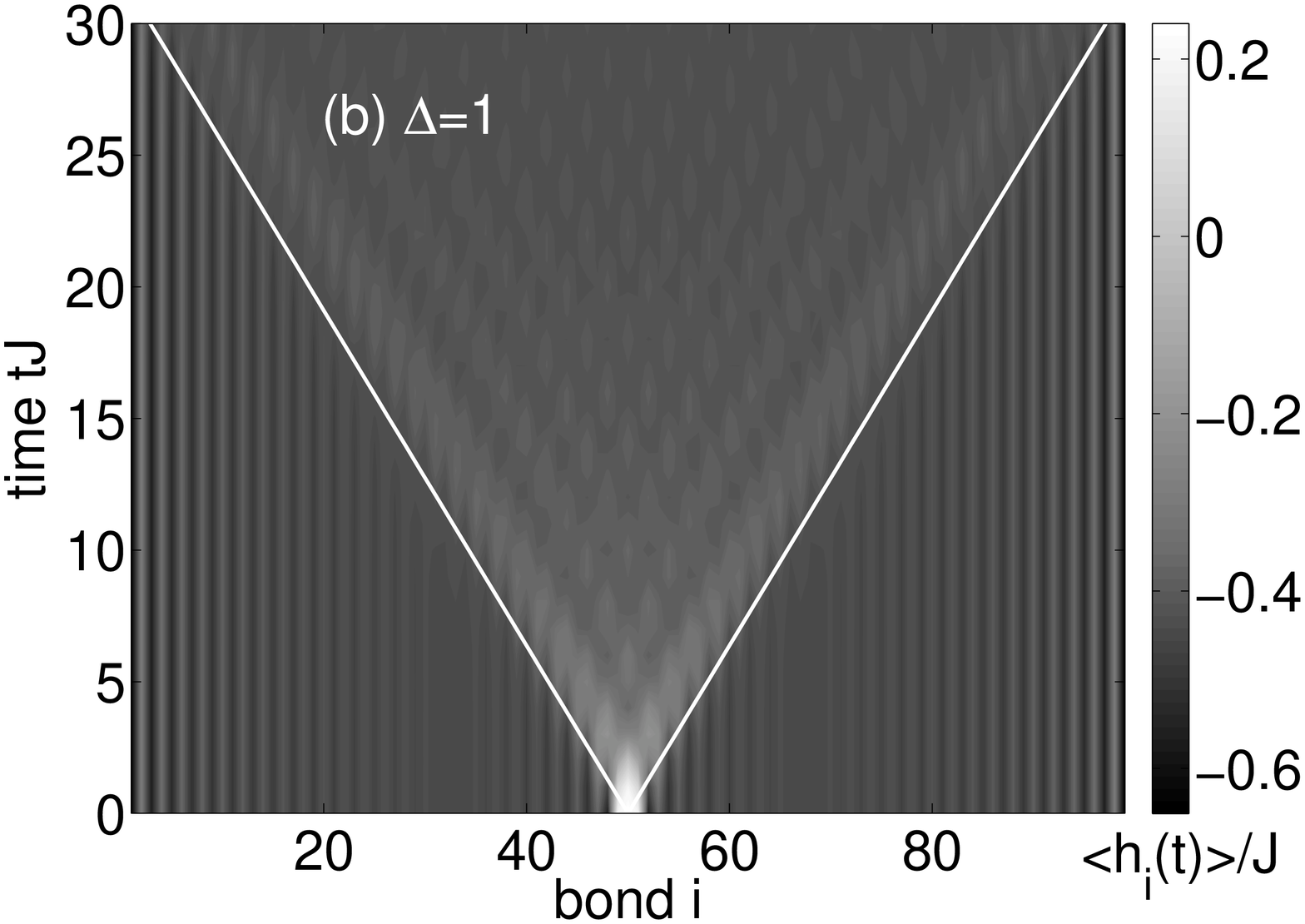}
   \includegraphics[width=0.3\textwidth]{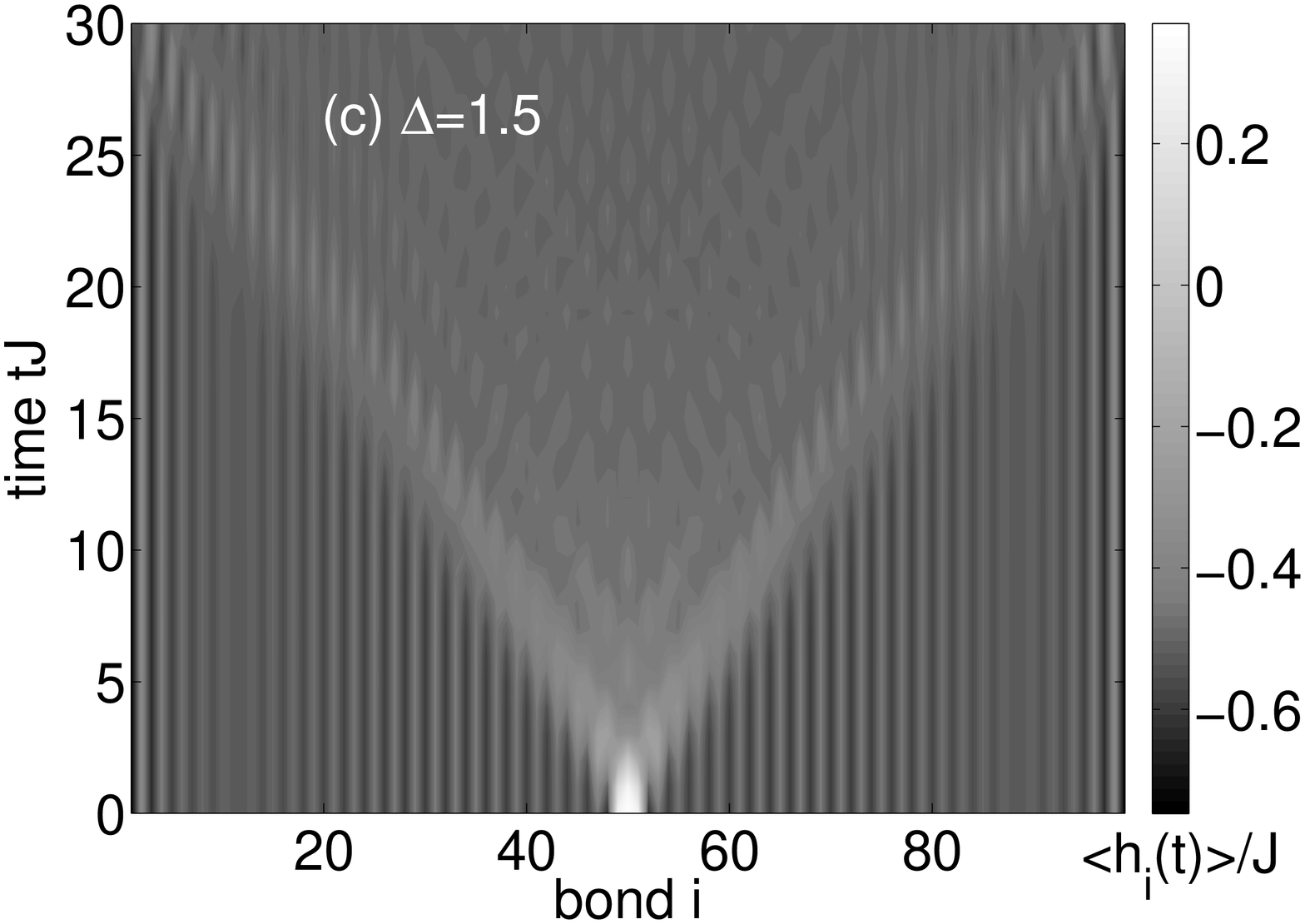}
  \caption{Time evolution of the bond energy distribution starting from initial states with $b=1$ from Fig.~\ref{fig:xxzinit} for (a) $\Delta=0.5$, (b) $\Delta=1$ and (c) $\Delta=1.5$. Despite the different ground state phases, for the selected values of the exchange anisotropy $\Delta$, main features of the dynamics such as two distinct rays extending from the edges of the perturbation are similar. The solid white lines for $\Delta=0.5$ and $\Delta=1$ indicate the propagation of a single excitations starting in the middle of the chain at time $t=0$ moving with the group velocity $v_g$ from Eq.~\eqref{eq:vg}. This is also the velocity in the outer rays.}
 \label{fig:timecont}
 \end{figure*}


\section{DMRG results for the $J_i$ quench}
\label{sec:4}

Now we turn to the numerical simulations. Using the adaptive time-dependent DMRG\cite{daley04,white04,vidal04,schollwoeck05,schollwoeck11} method we can access the real-time dynamics of initial bond energy distributions. Within this approach we can probe the microscopic dynamics including the time dependence of bond energies or the entanglement entropy starting from various initial states in an essentially exact manner without limitations in the range of parameters. We discuss the pure energy dynamics in the absence of spin currents induced by the $J_i$ quench in this section. We detail the construction of initial states and their specific features, then move on to the analysis of the time evolution of the bond energies. We calculate the spatial variance and the related quantity $J^*_E$ and discuss the emergent velocities of the energy dynamics. Within the numerical accuracy of our simulations we find a quadratic increase of $\sigma^2_E(t)$ in all cases studied. However, it seems that for a $J_i$ quench a large number of different velocities contribute as opposed to the Luttinger Liquid theory result, the latter valid at low energies.
Our study of the energy current during the time evolution and the time evolution of the expectation value $\langle J_E^*(t)\rangle$, defined in Eq.~\eqref{eq:Jstar}, gives additional insights into short-time dynamics and further validates the conclusion of ballistic energy dynamics.

\subsection{Initial states}
\label{sec:4a}
Let us first describe the typical shape of initial states induced by 
a $J_i$ quench on a few bonds in the middle of the spin chain. To be specific, in the Hamiltonian Eq.~\eqref{eq:hp}
we set
\begin{equation}
J_i=\left\lbrace\begin{array}{rcl}
 J&			& i<L/2-b 			\\ 
-J&\mbox{for}	& L/2-b\leq i \leq L/2+b    \\
 J&			& i>L/2+b \end{array}\right.\,,
\label{eq:block}
\end{equation}
which provides us with initial states with an inhomogeneous energy density profile with a width of $2b$ of the ferromagnetic region. Outside this ferromagnetic region we obtain antiferromagnetic nearest-neighbor correlations. 

Figure~\ref{fig:xxzinit} shows the profile of the local energy density of $XXZ$-chains with $L=100$ sites with (a) $\Delta=0.5$, (b) $\Delta=1$ and (c) $\Delta=1.5$, induced by a sign change of $J_i$ on $b=1,3,5$ bonds [compare Eq.~\eqref{eq:block}], 
obtained using DMRG with $m=200$ states exploiting the $U(1)$ symmetry to ensure 
zero global magnetization $S^z_{\mathrm{tot}}=\sum_i \langle S_i^z\rangle=0$ and, in consequence, $\langle S^z_i\rangle =0$. In all cases shown in Fig.~\ref{fig:xxzinit}, the system forms a region with ferromagnetic nearest-neighbor correlations in the middle of the chain. Note that for $\Delta\not= 1$, $\langle h_i\rangle$
is the sum of the nearest-neighbor transverse and longitudinal spin correlations, the latter weighted with $\Delta$.
In the regions with antiferromagnetic $J_i>0$, the local energy density oscillates, 
reflecting the antiferromagnetic nearest-neighbor correlations.
Figure~\ref{fig:energies} shows the energy difference $\delta E$. 
As a function of $b$, the energy difference $\delta E$ increases linearly once the smallest possible ferromagnetic region has been established. 
The minimum energy difference $\delta E=E_{\mathrm{init}}-E_0$ is of the order of $2J$, i.e., initial states that are only weak perturbations of 
the respective ground state cannot be generated using a $J_i$ quench.

At the isotropic point $\Delta=1$, we can explain the dependence of the initial state on the width $b$ in a transparent manner.
The ground state energy per site for the antiferromagnetic ground state is known from the Bethe Ansatz to be $\lim_{L\rightarrow\infty}E_0(L)/L= -\mbox{ln}(2)+1/4$,\cite{hulthen38} 
while for the ferromagnetic ground state $E_0/(L-1)=1/4$ excluding the boundary sites which gives rise to a very small system-size dependence. By growing the ferromagnetic region symmetrically with respect to the center of the chain and taking $E_0(L)$ from the unperturbed ground state with open boundaries, we obtain states with an 
energy that increases as 
\begin{equation}
\delta E(b)=(2*b-1)\cdot(E_0/(L-1)-0.25)+\delta E^0\,, \label{eq:dEexact}
\end{equation}
for our finite system size ($\delta E^0$ is simply an off-set). Equation~\eqref{eq:dEexact} exactly reproduces the data for $\Delta=1$ shown in Fig.~\ref{fig:energies}.
  
Furthermore, at $\Delta=1$, the total spin 
\begin{equation}
S_{\mathrm{tot}}^2=\sum_i \vec{S}_i\cdot \sum_j \vec{S}_j
\end{equation}
is a conserved quantity. Since the ground state calculation only respects the conservation of magnetization ($S^z_{\mathrm{tot}}=0$) we obtain a hierarchy of states 
with $S>0$. 
This can be easily understood by considering the block structure of the initial state. Taking, e.g., a total of $L=100$ 
spins and assuming a ferromagnetic region of only two spins (i.e., $b=1$), 
the two ferromagnetic spins are fully polarized with a total spin of $S=1$ while each of the 
antiferromagnetic blocks have 49 spins and therefore a total spin of $S=1/2$.
Thus, the total spin of the whole chain is $S_{\mathrm{tot}}=2$. Increasing the width of the ferromagnetic region by one, i.e., to $b=2$, 
we have $S=2$ in the middle and the antiferromagnetic blocks are of even length, which both have $S=0$ in their ground state. 
This pattern repeats itself upon increasing the length $2b$ of the ferromagnetic region.

\subsection{Time evolution of bond-energies after a $J_i$ quench}
\label{sec:4b}
Now we focus on the time-evolution of the local energy density induced by the aforementioned perturbation. At time $t=0^+$, we set all $J_i=J$ and then evolve under the dynamics of Eq.~\eqref{eq:xxz}.
The DMRG simulations are carried out using a Krylov-space based algorithm\cite{park86,hochbruck97} with a time-step of typically $0.25J$ and by enforcing
a fixed discarded weight. We restrict the discussion to times smaller than the time needed for the fastest excitation to reach the boundary. 

\subsubsection{$J_i$ quench: Qualitative features}

Figure~\ref{fig:timecont} shows the  time evolution of the bond energies $\langle h_i(t)\rangle $ as a contour plot for $\Delta=0.5,1,1.5$ at $b=1$. Despite the different ground states for the selected 
values of anisotropy, all features of the dynamics such as two distinct rays starting at the edges of the block of ferromagnetic correlations, are similar. 
The solid white lines for $\Delta=0.5$ and $\Delta=1$ indicate an excitation spreading out from the center of the ferromagnetic region with the group velocity given by Eq.~\eqref{eq:vg}
(these lines are parallel to the outer rays visible in the figure, i.e., the fastest propagating particles). 
Note that Eq.~\eqref{eq:vg} holds only in the gapless phase ($|\Delta|\leq1$). Besides the outer rays that define a light cone structrue, Fig.~\ref{fig:timecont} unveils the
presence of more such rays inside the light-cone. Since our particular initial states have a sharp edge in real space, there ought to be many exciations with different momenta $k$
contributing to the expansion.

\subsubsection{$J_i$ quench: Spatial variance}

\begin{figure}[tb]
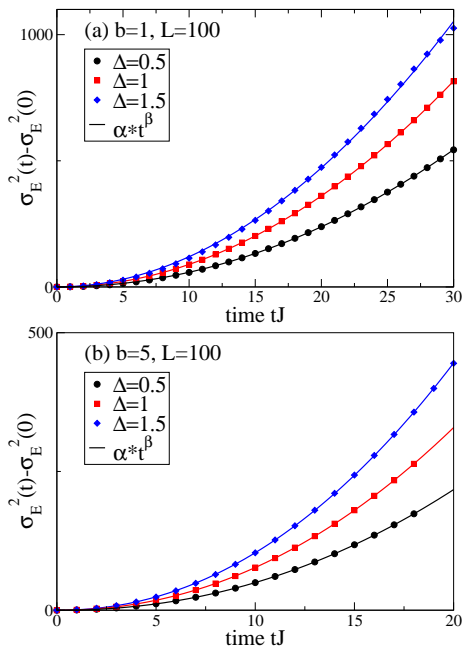

  \includegraphics[width=0.7\columnwidth]{fig5a.eps}
   \includegraphics[width=0.7\columnwidth]{fig5b.eps}
  \caption{(color online) Spatial variance of the evolving energy distribution for (a) $b=1$, (b) $b=5$ and $\Delta=0.5,1,1.5$. Fitting a power-law (straight lines) to $\sigma^2_E(t)-\sigma^2_E(0)=\alpha t^{\beta}$ yields a quadratic increase with sufficient accuracy, classifying the dynamics as ballistic. For instance we find $\alpha=0.53,\ \beta=2.03$ for $\Delta=0.5$ and $b=1$ [black circles in (a)]. We do not find any qualitative difference between the massless ($|\Delta|\leq1$) and the massive ($\Delta>1$) phase. The deviations between the fit and the tDMRG data in the $\Delta=1.5$ curves at the largest times simulated are due to boundaries.}
 \label{fig:fits}
 \end{figure}

Our main evidence for ballistic dynamics in both phases is based on the analysis of the spatial variance, shown in Fig.~\ref{fig:fits}. 
Fitting a power-law (straight lines) to the data, i.e., 
$\sigma^2_E(t)-\sigma^2_E(0)=\alpha t^{\beta}$ yields 
a quadratic increase with $\beta \approx 2$, classifying the dynamics as ballistic. 

In order to estimate uncertainties in the fitting parameter $\alpha$, we compare this to the results of fitting a pure parabola $\sigma^2_E(t)-\sigma^2_E(0)=V_E^2\, t^2$ 
to the data.
Typically, $V_E^2$ deviates from $\alpha$ by about $10\%$ while the exponent of the power-law fit is usually different from $2$ by $5\%$. 
As an example, for $\Delta=0.5$, $b=1$ we obtain $\beta=2.03$ and $\alpha=0.53$ vs. $V_E^2=0.6J^2$. 
The main reason for the deviation of $\beta$ from $2$ is, in fact, that the short time dynamics is not well described by a power law at all
over a $b$-dependent time-window. We shall see later, in Sec.~\ref{sec:jstar}, that the ballistic dynamics sets in only after the block of ferromagnetically
correlated bonds has fully 'melted'.
Indeed, by excluding several time steps at the beginning of the evolution from the power-law fit, we observe that $\beta \rightarrow 2$ and $\alpha\rightarrow V_E^2$. 
Therefore, we will present results for $V_E^2$, obtained by fitting $\sigma^2_E(t)-\sigma^2_E(0)=V_E^2 t^2$ to our tDMRG data.

\subsubsection{Exploiting $SU(2)$ symmetry at $\Delta=1$ for the $J_i$ quench}
\label{sec:4c}
\begin{figure}[tb]
  \includegraphics[width=0.7\columnwidth]{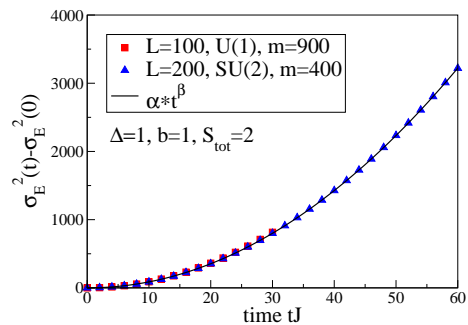}
  \caption{(color online) Long-time evolution exploiting the conservation of total spin $S_{\mathrm{tot}}$ at $\Delta=1$ for $L=200$ sites using an initial state with
$b=1$. For comparison we plot the result for $L=100$ sites using only $U(1)$ symmetry. Fixing the discarded weight to $10^{-4}$ we need less than half the number of states. Furthermore, we find that the spatial variance is very robust against finite-size effects.}
 \label{fig:SU2}
 \end{figure}

Before proceeding to the discussion of the expansion velocity $V_E^2$, we wish to 
discuss the long-time limit, which can be accessed in the case of $\Delta=1$. Since our perturbation is proportional to the operators for the local energy density, global symmetries of the unperturbed Hamiltonian are respected by the initial states
of the type Eq.~\eqref{eq:block}. Therefore, at $\Delta=1$, we can exploit
 the conservation of total spin $S$, a non-Abelian symmetry.
This can be used to push the simulations to much longer times,
since we can perform the time evolution in an $SU(2)$ invariant basis.\cite{mcculloch02} The number of states needed to ensure a given accuracy is reduced substantially compared to a simulation that only respects $U(1)$ symmetry. Therefore, we can work with larger system sizes and
 study the long-time dynamics of the energy density. As we can reach longer times, we can also analyze and discuss finite-size effects for $\Delta=1$ here. Figure~\ref{fig:SU2} shows our result for the time evolution respecting $SU(2)$ symmetry (blue triangles) for a system of $L=200$ sites and $\Delta=1$, $b=1$ compared to the result from Fig.~\ref{fig:fits} for $L=100$ sites (red squares). We still find a quadratic increase of $\sigma^2_E(t)$ and thus ballistic dynamics for times up to $t\sim 60/J$ and in addition, the prefactor does not depend on the system size. Both simulations were carried out keeping the discarded weight below $10^{-4}$ which requires at most $m=900$ states using only $U(1)$ symmetry on $L=100$ sites versus a maximum of $m=400$ using $SU(2)$ for $L=200$ sites.

\subsubsection{Expansion velocity}

\begin{figure}[tb]
  \includegraphics[width=0.7\columnwidth]{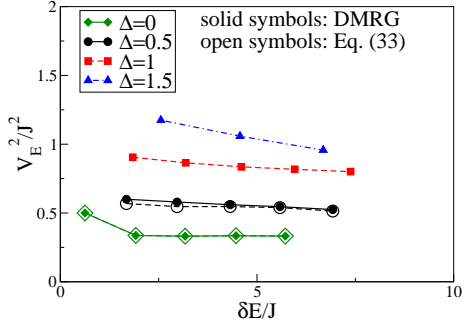}
  \caption{(color online) Prefactors $V_E^2$ of the fits $\sigma^2_E-\sigma^2_E(0)=V_E^2 t^2$ as a function of $\delta E$ for $\Delta=0,0.5,1,1.5$ and $J_i$ quenches with $b=1,2,3,4,5$ (for $\Delta=1.5$, we show $b=1,2,3$ only). For $\Delta>0$, $V_E^2$ decreases slightly with $b$ while $V_E^2<v_g^2$. At $\Delta=0$, $V_E^2$ is roughly constant for $b>2$.}
 \label{fig:VE}
 \end{figure}

The results for $V_E^2$ are collected in Fig.~\ref{fig:VE} and plotted as a function of $\delta E$ for $\Delta=0,0.5,1,1.5$. 
In the non-interacting case, $\Delta=0$, $V_E^2$ is constant for $b\geq 2$, while at $b=1$ (the smallest possible $\delta E$), $V_E^2=0.5J^2$.
For all $\Delta>0,\ V_E^2$ slightly decreases with $\delta E$ and $V_E^2$ is much smaller than $v_g^2$ given by Eq.~\eqref{eq:vg}, suggesting that indeed, many velocities contribute during the
expansion of the energy wave-packet.

Intuitively, one might associate the decrease of $V_E^2$, which is a measure of the average velocity of propagating excitations
contributing to the expansion, to band curvature: The higher $\delta E$, the more excitations with velocities smaller than $v_g$ 
are expected to factor in. 

\begin{figure}[tb]
  \includegraphics[width=0.7\columnwidth]{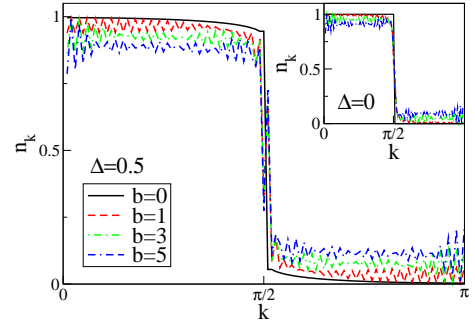}
  \caption{(color online) MDF of the initial states generated by a $J_i$ quench at $\Delta=0.5$ and $\Delta=0$ (inset), with $b=1,3,5$. For comparison we include the MDF of the groundstate (solid black line).}
 \label{fig:mdf}
 \end{figure}

It is instructive to consider the non-interacting limit first by comparing the numerical results
obtained from a time-evolution with exact diagonalization to the analytical (and also exact result)
from Eq.~\eqref{eq:v2}. To that end we need to compute the MDF [see Eq.~\eqref{eq:mdf}] of the initial state.
Our results for $\Delta=0$, which are shown in Fig.~\ref{fig:mdf}, unveil
a peculiar property: The $J_i$ quench always induces changes at all $k$, i.e., the system is not just weakly perturbed in the vicinity of $k_F$. This is not surprising since our initial states have sharp edges in real-space (compare Fig.~\ref{fig:xxzinit}). Moreover, the $J_i$ quench changes the MDF in such a way that $\delta n_k(b)=n_k^{\mathrm{init}}(b)-n_k$ is point-symmetric with respect to $k_F=\pi/2$ , where $k_F$ is the Fermi wave-vector. 
As Fig.~\ref{fig:VE} shows, $V_E^2$ as extracted from fits to $\delta \sigma_E^2$ (solid symbols) 
and $V_E^2$ from Eq.~\eqref{eq:v2} [open symbols] perfectly agree with each other, as expected.

The MDF of initial states for the {\it interacting} systems are also such that $\delta n_k\not=0 $ at all momenta
and we may therefore conclude that the observation $V_E<v_g$ is due to the fact that
the $J_i$ quench induces many excitations
with velocities smaller than $v_g$ (compare the data shown for $\Delta=0.5$ shown in Fig.~\ref{fig:mdf}). Of course, Eq.~\eqref{eq:v2} is not directly applicable to the interacting case
since, first, it does not account for the correct eigenstates at $\Delta\not=0$ and second, in general, $\langle h_i \rangle \not=\langle J(S^+_i S^-_{i+1}+h.c.)/2\rangle $.
Nevertheless, by numerically calculating $\delta n_k$ for the interacting system and by using the 
renormalized velocity in Eq.~\eqref{eq:v2} instead of $J$ [i.e., $J\rightarrow v_g(\Delta)$], we obtain an estimate for $V_E^2$ from
\begin{equation}
V_E^2\approx\frac{v_g^2}{\delta E} \sum_k \cos(k)\sin^2(k) \delta n_k \,.\label{eq:v2delta}
\end{equation}
This reproduces the qualitative trend of the tDMRG results for $V^2_E$ as we exemplify for $\Delta=0.5$ in Fig.~\ref{fig:VE}.

To summarize, the overall picture for the time evolution of the bond energies after a $J_i$ quench is: 
Energy propagates ballistically with an expansion velocity $V_E$ that is approximately given by Eq.~\eqref{eq:v2delta}.
 Combined with the observation that on a finite system, a $J_i$ quench induces changes in the MDF
at all momenta $k$, we conclude that many excitations contribute to the wave-packet dynamics, resulting in $V_E < v_g$, both in
the non-interacting and in the interacting case.


\subsection{Energy currents}
\label{sec:jstar}
\begin{figure}[tb]
  \includegraphics[width=0.7\columnwidth]{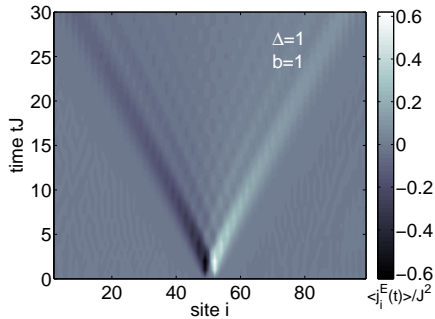}
  \caption{(color online) Real-time evolution of the local energy current Eq.~\eqref{eq:jenergy} at $\Delta=1$ for a $J_i$ quench with $b=1$.}
 \label{fig:current}
 \end{figure}

\begin{figure}[tb]
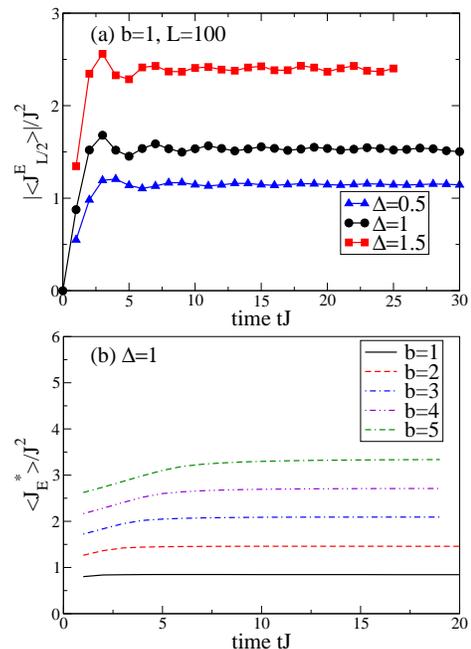

  \includegraphics[width=0.7\columnwidth]{fig10a.eps}
  \includegraphics[width=0.7\columnwidth]{fig10b.eps}
  \caption{(color online) (a) Absolute value of the current in each half of the system. A constant value is reached after $t\approx 5/J$. (b) The quantity $\langle J^*_E(t)\rangle$ from Eq.~\eqref{eq:Jstar} derived from a pure quadratic increase of the spatial variance for $\Delta=1$ and $b=1,2,3,4,5$. This quantity is constant, as expected from the discussion in Sec.~\ref{sec:jstar}, except
for the initial transient dynamics at $t< b/J$.}
 \label{fig:current2}
 \end{figure}

To conclude the discussion of the $J_i$ quenches we present our results for the local energy currents at $\Delta=1$ in Fig.~\ref{fig:current}. 
By comparison with Fig.~\ref{fig:timecont}(b), we see that the local current is the strongest in the vicinity of the wave packet. 
The energy current in each half of the system becomes a constant after a few time steps, i.e., $J^E_{L/2}:=\sum_{i=1}^{L/2-1} j^E_i$ reaches a constant value. We plot the absolute value of $\langle J^E_{L/2}\rangle$ for $\Delta=0.5,1,1.5$ for $b=1$ in Fig.~\ref{fig:current2}(a). The qualitative behavior 
is  independent of $\Delta$: As soon as the initial perturbation has split up into two wave packets, 
we have prepared each half
of the chain in a state with a constant, global current $\langle J^E_{L/2}\rangle$=const.
 For a system with periodic boundary conditions, 
the total current $J_E=\sum_ij_i^E$ is a conserved quantity.\cite{zotos97} 
Since the effect of boundaries only factors in once these are reached by the fastest excitations, we directly probe the
conservation of a global current with our set-up, after some initial transient dynamics. Therefore, we can link
the phenomenological observation of ballistic wave-packet dynamics to the existence of a conservation law in the system. 

While the currents $\langle J^E_{L/2}\rangle$ clearly undergo some transient dynamics [see Fig.~\ref{fig:current2}(a)],
we have derived a quantity in Sec.~\ref{sec:model}, called $J^*_E$, whose expectation value is stationary if $\sigma_E^2 \sim t^2$.
We now numerically evaluate $\langle J^*_E(t)\rangle$ from Eq.~\eqref{eq:Jstar} which provides an independent probe of ballistic dynamics. 
Figure~\ref{fig:current2}(b) shows our results for $\Delta=1$ and $J_i$ quenches with $b=1,2,3,4,5$. 
It turns out that $\langle J^*_E(t)\rangle$ is indeed constant at sufficiently large times, consistent with the observation of $\delta \sigma_E^2 \sim t^2$.
In Sec.~\ref{sec:4b}, we have noted that $\delta \sigma_E^2 \not\sim t^2$ at short times $t\lesssim b/J$. 
 This renders $\langle J_E^*(t)\rangle$ a time-dependent quantity over the same time window: Clearly, 
the time window over which $\langle J_E^*(t)\rangle \not=$ const, depends on $b$ [see Fig.~\ref{fig:current2}(b)], which suggests that
the deviation of ballistic dynamics is associated to the 'melting' process of the region with ferromagnetic correlations.
We have carefully checked that these observations are robust against errors in the calculation of time derivatives in Eq.~\eqref{eq:Jstar} induced 
by the finite time step. Since $\langle J_E^*(t)\rangle$ is time-dependent (at least at short times), we conclude that $J_E^*$ is not a conserved quantity in the interacting case.
Finally, within our numerical accuracy and as an additional consistency check, we find that
 $\langle J^*_E\rangle/\delta E=\alpha$ in the stationary state as expected from the discussion in Sec.~\ref{sec:currents}.


To summarize, $\langle J^*_E(t)\rangle$=const whenever $\delta \sigma_E^2\sim t^2$ but $\langle J^*_E\rangle $ is very sensitive to the initial transient dynamics in the energy dynamics and becomes constant after a time $\approx bJ$.
Furthermore, our set-up serves to prepare each half of the system in a state with a finite global energy current $\langle J_{L/2}^E\rangle$ that, after some transient dynamics, does not decay since the global energy current operator is a conserved quantity.

 \section{Coupled spin and energy dynamics}
 \label{sec:5}
 After focussing on the energy dynamics in the absence of spin-/particle currents we now revisit the case of spin dynamics starting from states with $\langle S^z_i(t=0)\rangle\not= 0$. Thus, during the time evolution, the local spin and energy currents are both non-zero.
In Ref.~\onlinecite{langer09}, the dynamics of the magnetization was studied, where the inhomogeneous spin density profile was induced by a Gaussian magnetic field in the initial state. We take the initial state to be the ground state of Eq.~\eqref{eq:gaussB0} in the sector with zero global magnetization, i.e., $S^z_{\mathrm{tot}}=\sum_i\langle S^z_i\rangle=0$. Such a perturbation naturally also results in an inhomogeneous energy density in the initial state, which is coupled to the spin dynamics during the time evolution. \cite{jesenko11} 

\subsection{Massless phase}
\begin{figure}[tb]
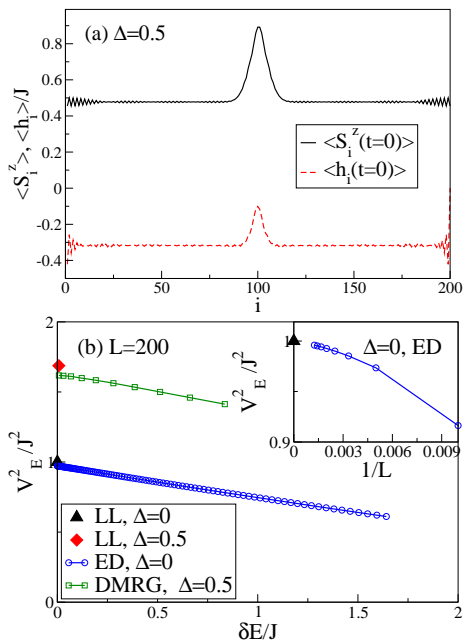

  \includegraphics[width=0.7\columnwidth]{fig11a.eps}
  \includegraphics[width=0.7\columnwidth]{fig11b.eps}
  \caption{(color online) (a) Magnetization (solid black line) and energy density (dashed red line) in the initial state, for a $B_0$ quench with $B_0=J$ and $\sigma_0=5$ for $\Delta=0.5$ on a lattice of $L=200$ sites. (b) Prefactor $V^2_E$ of $\delta \sigma_E^2(t)=V^2_Et^2 $ for the energy dynamics after a $B_0$ quench in the massless phase of the XXZ chain, compared to the group velocity [Eq.~\eqref{eq:vg}] for $\Delta=0,0.5$ and $L=200$. On this system size and in the limit of small perturbations, $V^2_E$ is approximately $5\%$ smaller than the prediction from the Luttinger Liquid theory for both $\Delta$. For $\Delta=0$, finite-size scaling of $V_E^2(\delta E\rightarrow 0)$ using $L=100,200, ..., 800$ yields $V^2_E\rightarrow v_g^2$ as shown in the inset.}
 \label{fig:Bgauss}
 \end{figure}

In Fig.~\ref{fig:Bgauss}(a) we compare the initial magnetization (black solid line) and the local bond energies (dashed red line) induced by a Gaussian magnetic field with $B_0=J$ and $\sigma_0=5$ at $\Delta=0.5$ finding qualitatively the same pattern: Both the spin and the energy
density follow the shape of the magnetic field, resulting in a smooth perturbation with small oscillations in the background away from the wave packet.

For the time evolution of the bond energies at $0<\Delta\leq1$, we perform an analysis of their spatial variance analogous to the discussion of the $J_i$ quench, 
finding ballistic dynamics in the massless phase. Since with a $B_0$ quench, initial states with very small $\delta E$ can be produced, we next connect our numerical results
to the predictions of LL theory, valid in the limit $\delta E \ll J$ (compare Sec.~\ref{sec:ll}).
 
Since we enforce zero global magnetization, we draw magnetization from the background into the peak.\cite{kollath05b} Therefore, one has to carefully estimate the contributions to $\delta E$ 
that do not contribute to the time-dependence of bond energies yet change the background density $n_{bg}$. The latter, in turn, affects the expected group velocity and we thus
expect to recover the LL result derived for the half-filled case, i.e., propagation with $v_g$ from Eq.~\eqref{eq:vg}, in the limit of large systems where $n_{bg}\to 1/2$. 
Furthermore, $B_0$ quenches induce $2k_F$-oscillations in the spin and energy-density.\cite{langer09}
To account for this we use coarse graining, i.e., averaging the energy density over neighboring sites, and we take the sum only over the area of the peak when estimating
$\delta E$: 
We obtain $\delta E^{\mathrm{peak}}:=\sum_{L/2-x}^{L/2+x}(\langle h_i\rangle-\langle h_i\rangle_0)$ where $\langle h_i\rangle_0$ denotes the ground state expectation value. From this quantity we calculate the velocity via $V^2_E\rightarrow V_E^2\cdot\delta E/\delta E^{\mathrm{peak}}$, which is shown in Fig.~\ref{fig:Bgauss}(b). Note that while $\delta E_{\mathrm{peak}}$ is the correct normalization to obtain the correct velocities, we label our initial states via $\delta E$. At $\Delta=0$ (blue circles), $V^2_E$ decreases linearly as a function of $\delta E$. Next we compare the result from the 
low-energy theory from Sec.~\ref{sec:ll} (solid symbols at $\delta E=0$) to our tDMRG data. For both $\Delta=0$ and $\Delta=0.5$, $V^2_E$ for $L=200$ sites is approximately $5\%$ smaller than $v_g^2$ from Eq.~\eqref{eq:vg},
which is mainly due to the deviation of the background density from half filling. 
While it is hard to get results for larger systems than $L\sim 200$ in the interacting case, we can solve the $\Delta=0$ case numerically exactly 
in terms of free spinless fermions, allowing us to go to sufficiently large $L$ to observe $V^2_E(L)\rightarrow v_g^2$ as $L\rightarrow \infty$.
 The inset of Fig.~\ref{fig:Bgauss}(b) shows the finite-size scaling of $V^2_E(L)$ for $\Delta=0$ using $L=100,200, ..., 800$ 
which yields $V^2_E\rightarrow v_g^2$ in the limit $L\to\infty$, taking first $\delta E\rightarrow 0$ for each system size. 
We thus, in principle, have numerical access to the dynamics in the low energy limit well described by Luttinger Liquid theory using a $B_0$ quench.

\subsection{Massive phase}
In Ref.~\onlinecite{langer09} examples of a linear increase of the spatial variance of the magnetization $\sigma_S^2(t)$, defined in Eq.~\eqref{eq:sigmas}, were found in the massive phase, which were
interpreted as an indication of 
diffusive dynamics. 
We now demonstrate that while the spin dynamics may behave diffusively, i.e., $\delta\sigma_S^2 \sim t$ over a certain time window, the energy
dynamics in the same quench is still ballistic, i.e., $\delta\sigma_E^2 \sim t^2$.

 \begin{figure}[t]
 \includegraphics[width=0.7\columnwidth]{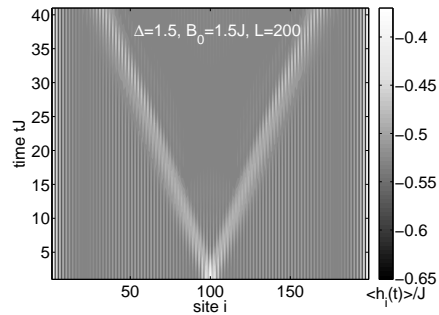}
 \caption{(color online) Time dependent bond energies for the dynamics induced by a $B_0$ quench with $B_0=1.5J,\ \sigma_0=5$ on a chain of $L=200$ sites at $ \Delta=1.5$: In this 
case, both local spin and local energy densities are perturbed and the corresponding local currents are non-zero. }
\label{fig:combcont}
\end{figure}

\begin{figure}
  \includegraphics[width=0.7\columnwidth]{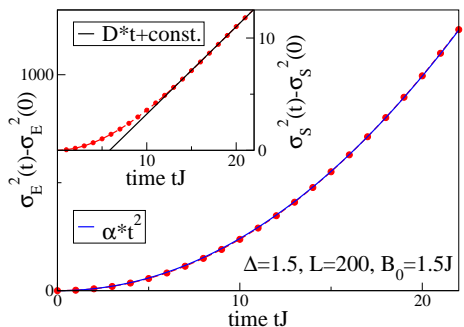}
  \caption{(color online) Spatial variance of the energy density (main panel) and the spin density (inset), induced by a $B_0$ quench with $B_0/J=1.5$ and $\sigma_0=5$ [compare Eq.~\eqref{eq:gaussB0}] at $\Delta=1.5$: In this 
case, both local spin and local energy densities are non-zero during the time-evolution. The inset was reproduced from Ref.~\onlinecite{langer09}.} 
 \label{fig:coupled}
 \end{figure}

In Fig.~\ref{fig:combcont}, we show the full time evolution of the bond energies for a Gaussian magnetic field with $B_0=1.5J$ and $\sigma_0=5$ on a chain of $L=200$ sites at $\Delta=1.5$. It consists of two rays propagating with opposite velocities. In Fig.~\ref{fig:coupled}, we 
compare the spatial variance of the magnetization $\sigma^2_S(t)$ to the one of the bond energies $\sigma^2_E(t)$ calculated in the same time evolved state. 
The main panel of Fig.~\ref{fig:coupled} shows $\sigma_E^2(t)-\sigma_E^2(0)$ which is very well described by a power-law fit with an exponent $\beta=1.98$ on the accessible time scales. 
The inset of Fig.~\ref{fig:coupled} displays the data for $\delta \sigma_S^2(t)=\sigma_S^2(t)-\sigma^2_S(0)$ taken from Ref.~\onlinecite{langer09}. The spatial variance 
of the energy density is quadratic in time, even at times $t\gtrsim 12/J$ where the spatial variance of the magnetization increases only linearly. 
This example reflects the qualitative difference between spin and energy transport in the massive phase of the XXZ model 
at zero global magnetization: The conservation of the global energy current is consistent with the observation of ballistically propagating 
 energy wave-packets while spin 
clearly does not propagate ballistically. Our result, obtained in the non-equilibrium case with a zero-temperature background density, is consistent with the
picture established from both linear-response theory\cite{prelovsek04,steinigeweg09} and steady-state simulations.\cite{prosen09,jesenko11,znidaric11}

Very recently, Jesenko and $\check{\mbox{Z}}$nidari$\check{\mbox{c}}$ have also studied the time evolution of
spin and energy densities induced by a $B_0$ quench.\cite{jesenko11} They concentrate 
their analysis on the velocity of the fastest wave-fronts, 
contrasting energy against spin dynamics. Based on the presence of these rays of fast propagating particles,
they claim that the wave-packet dynamics still has ballistic features. However, their analysis neglects the influence
of slower excitations that also contribute to the dynamics of the
wave packet, which is captured by the variance, and it ignores the
decay of the intensity in the outer rays that we typically observe whenever $\delta \sigma_S^2\sim t$.\cite{langer09} The latter is, if at all, weak in
a ballistic expansion characterized by $\delta\sigma_S^2\sim t^2$. 
Therefore, while the analysis of Ref.~\onlinecite{jesenko11} unveils interesting details of the time evolution of densities during a $B_0$ quench,
we maintain that the variance is a useful quantity to identify candidate parameter sets for spin diffusion in, e.g., the non-equilibrium regime.
Final proof of diffusive behavior then needs to be established by either demonstrating the validity of the diffusion equation or by computing correlation functions, see, e.g., Ref. \onlinecite{steinigeweg09,znidaric11}. For instance, in Ref.~\onlinecite{jesenko11}, Jesenko and $\check{\mbox{Z}}$nidari$\check{\mbox{c}}$ analyze the steady-state currents in the $\Delta>1$ regime at finite temperature and obtain diffusive behavior.
  
\section{Summary}
\label{sec:sum}
We studied the real-time energy dynamics in XXZ spin-1/2 chains at zero temperature in two different scenarios. First, we investigated the energy dynamics in the 
absence of spin currents induced by a local sign change in the exchange interactions.
The spatial variance behaves as $\delta\sigma_E^2(t) \propto t^2$ for all $\Delta$, 
consistent with ballistic dynamics. In the gapless regime, the velocity of the fastest excitation present in the dynamics is the group velocity $v_g$ of spinons, yet our particular quench also involves excitations with much smaller 
velocities resulting in expansion velocities $V_E<v_g$. 
Furthermore, the ballistic dynamics can be related to properties of energy currents. While the total current vanishes in our set-up, i.e., $\langle J_E\rangle:=\sum_i\langle j_i^E\rangle=0$,
the current in each half of the chain $\langle J_{L/2}^E\rangle>0$ takes a constant value, after some transient dynamics. Therefore,
in each half of the system, we prepared a state with a conserved global current, allowing us to make a
direct connection to the predictions of linear-response theory where the existence of ballistic dynamics is directly
linked to conservation laws that prohibit currents from decaying.\cite{zotos97} 
Moreover, we identified an observable $J_E^*$ built from local currents whose expectation value $\langle J_E^*(t)\rangle $ is time independent if 
$\delta \sigma^2_E \propto t^2$ and vice versa. This carries over to other types of transport as well
and, in fact, the analysis of the time-dependence of $\langle J_E^*\rangle$ can be used as an independent means
to identify ballistic regimes, or to unveil the absence thereof. 

In the second part, we studied the energy dynamics induced by quenching a Gaussian magnetic field, with two main results.
These quenches allow us to access the regime of weakly perturbed initial states and in that limit, we recover the 
predictions from Luttinger Liquid theory for the wave-packet dynamics: Their variance simply grows as $\delta \sigma_E^2 =v_g^2 t^2$. 
 In the massive phase a very interesting phenomenon occurs, since the energy dynamics is ballistic on time scales over which the spin dynamics behaves diffusively 
although both are driven by the same perturbation. This resembles the picture established from linear-response theory,\cite{zotos97} there applied to the finite-temperature case, 
in the non-equilibrium setup studied here. While our numerical results cover spin chains on real space lattices and initial states far from equilibrium, the extension of our work to a finite temperature of the background will be crucial to tackle the most important open questions.

\acknowledgements
We thank W. Brenig, S. Kehrein, A. Kolezhuk, and R. Noack for very helpful discussions.
S.L. and F.H-M. acknowledge support from the {\it Deutsche Forschungsgemeinschaft} through FOR 912, 
M.H acknowledges support by the SFB TR12 of the {\it Deutsche Forschungsgemeinschaft}, the Center for Nanoscience (CeNS) Munich, and the German Excellence Initiative via the Nanosystems Initiative Munich (NIM).
F.H-M. acknowledges the hospitality of the Institute for Nuclear Theory at the University of Washington, Seattle, where part of this research was carried out
during the INT program {\it Fermions from Cold Atoms to Neutron Stars: Benchmarking the Many-Body Problem}.

\appendix

\section{Entanglement growth}

Here we want to study the growth of entanglement across a junction separating regions in a spin chain with ferromagnetic correlations
from ones with antiferromagnetic ones. 

\begin{figure}[ht]  \includegraphics[width=0.7\columnwidth]{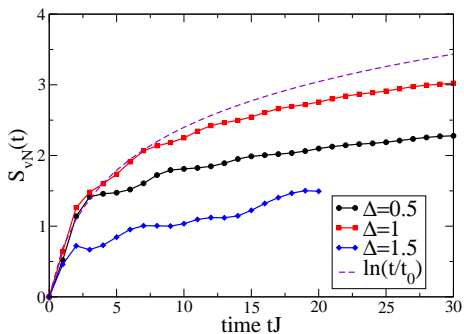}  
 \caption{(color online) Time dependence of the von-Neumann Entropy $S_{\mathrm{vN}}$ 
for a bipartition that cuts the system across the central bond during the time evolution starting from a ferromagnetic region coupled to an antiferromagnetic one at $\Delta=0.5,1,1.5$}
 \label{fig:Entropy}
 \end{figure}

To that end, we take initial states inspired by Ref.~\onlinecite{gobert05} where one half of the system has a positive and the other one a negative $J$.
 We obtain this configuration as a variation of $J_i$ quench choosing:
\begin{equation}
J_i=\left\lbrace\begin{array}{rcl}
 J&	& i<L/2 			\\ 
 0&	\mbox{for}	& i=L/2 			\\ 
-J& & i>L/2    \\
\end{array}\right.\,,
\label{eq:gblock}
\end{equation}
in Eq.~\eqref{eq:hp}.
We then perform the time evolution under the antiferromagnetic Hamiltonian [Eq.~\eqref{eq:xxz}]. As a measure of the entanglement we calculate the von Neumann entropy
\begin{equation}
S_{\mathrm{vN}}=-Tr(\rho_A ln \rho_A)
\end{equation}
of the reduced density matrix $\rho_A=Tr_B\rho$, where $\rho=|\psi(t)\rangle\langle\psi(t)$ and $|\psi(t)\rangle$ is the time-evolved wave function, for a bipartition in which we cut the chain into two halves of length $L/2$ across the central link. 
Our results are plotted in Fig.~\ref{fig:Entropy}. We observe that the von-Neumann entropy grows  at most logarithmically (purple dashed line), 
 in agreement with Ref.~\onlinecite{eisler09}. The overall largest values of $S_{\mathrm{vN}}(t)$ are found at the critical point $\Delta=1$ (red squares). 
This behavior is very similar to the observations made in Ref.~\onlinecite{gobert05} for spin dynamics starting from a state with all spins pointing up(down)
in the left(right) half.

\begin{thebibliography}{10}

\bibitem{zotos-review}
X.~Zotos and P.~Prelov{\v{s}}ek,
\newblock {\em Physics and Chemistry of Materials with Low-Dimensional Structures},
  (Kluwer Academic, Doodrecht, 2004), Chap. 11.

\bibitem{hm07a}
F.~Heidrich-Meisner, A.~Honecker, and W.~Brenig,
\newblock Eur. J. Phys. Special Topics {\bf 151}, 135 (2007).

\bibitem{rozhkov05}
A.V.~Rozhkov and A.L.~Chernyshev,
\newblock Phys. Rev. Lett. {\bf 94} 087201 (2005).

\bibitem{boulat07}
E.~Boulat, P.~Mehta, N.~Andrei, E.~Shimshoni and A.~Rosch,
\newblock Phys. Rev. B, {\bf 76} 214411 (2007).

\bibitem{bloch08}
I.~Bloch, J.~Dalibard, and W.~Zwerger,
\newblock Rev. Mod. Phys. {\bf 80}, 885 (2008).

\bibitem{polkovnikov11}
A.~Polkovnikov, K.~Sengupta, A.~Silva, and M.~Vengalattore,
\newblock Rev. Mod. Phys., {\bf 83} 863 (2011).

\bibitem{schneider11}
U.~Schneider, L.~Hackerm\"uller, J.~P. Ronzheimer, S.~Will, S.~Braun, T.~Best,
  I.~Bloch, E.~Demler, S.~Mandt, D.~Rasch, and A.~Rosch,
\newblock preprint arXiv:1005.3545  (unpublished).

\bibitem{medley11}
P.~Medley, D.~M. Weld, H.~Miyake, D.~E. Pritchard, and W.~Ketterle,
\newblock Phys. Ref. Lett. {\bf 106} 195301 (2011)

\bibitem{sommer11a}
A.~Sommer, M.~Ku, and M.~W. Zwierlein,
\newblock New J. Phys.{\bf 13}, 055009 (2011)

\bibitem{sommer11b}
A.~Sommer, M.~Ku, G.~Roati, and M.~W. Zwierlein,
\newblock Nature {\bf 472}, 201 (2011)

\bibitem{joseph11}
J.~Joseph, J.~E. Thomas, M.~Kulkarni, and A.~G. Abanov,
\newblock J.Stat.Mech.: Theory Exp. {\bf(2011)} P04007

\bibitem{rigol04}
M.~Rigol and A.~Muramatsu,
\newblock Phys. Rev. Lett. {\bf 93}, 230404 (2004).

\bibitem{rigol05b}
M.~Rigol and A.~Muramatsu,
\newblock Phys. Rev. Lett. {\bf 94}, 240403 (2005).

\bibitem{kollath05b}
C.~Kollath, U.~Schollw\"{o}ck, J.~von Delft, and W.~Zwerger,
\newblock Phys. Rev. A {\bf 71}, 053606 (2005).

\bibitem{hm08a}
F.~Heidrich-Meisner, M.~Rigol, A.~Muramatsu, A.~E. Feiguin, and E.~Dagotto,
\newblock Phys. Rev. A {\bf 78}, 013620 (2008).

\bibitem{hm09a}
F.~Heidrich-Meisner, S.~R. Manmana, M.~Rigol, A.~Muramatsu, A.~E. Feiguin, and
  E.~Dagotto,
\newblock Phys. Rev. A {\bf 80}, 041603(R) (2009).

\bibitem{karlsson11}
D.~Karlsson, M.~O. C.~Verdozzi, and K.~Capelle,
\newblock EPL {\bf 93}, 23003 (2011).

\bibitem{kajala11}
J.~Kajala, F.~Massel, and P.~T\"orm\"a,
\newblock Phys. Rev. Lett. {\bf 106}, 206401 (2011)
\bibitem{gobert05}
D.~Gobert, C.~Kollath, U.~Schollw\"ock, and G.~Sch\"utz,
\newblock Phys.\ Rev.\ E {\bf 71}, 036102 (2005).

\bibitem{langer09}
S.~Langer, F.~Heidrich-Meisner, J.~Gemmer, I.~McCulloch, and U.~Schollw\"ock,
\newblock Phys. Rev. B {\bf 79}, 214409 (2009).

\bibitem{eisler09}
V.~Eisler and I.~Peschel,
\newblock J. Stat. Mech.: Theor. Exp. (2009) P02011.

\bibitem{lancaster10a}
J.~Lancaster and A.~Mitra,
\newblock Phys. Rev. E {\bf 81}, 061134 (2010).

\bibitem{lancaster10b}
J.~Lancaster, E.~Gull, and A.~Mitra,
\newblock Phys. Rev. B {\bf 82}, 235124 (2010).

\bibitem{mossel10}
J.~Mossel, G.~Palacios, and J.-S. Caux,
\newblock J. Stat. Mech.: Theory Exp. (2010) L09001.

\bibitem{foster10}
M.~S.~Foster, E.~A.~Yuzbashyan, B.~L.~Altshuler
\newblock Phys. Rev. Lett. {\bf 105}, 135701 (2010) 

\bibitem{foster11}
M.~S.~Foster, T.~C.~Berkelbach, D.~R.~Reichman and E.~A.~Yuzbashyan
\newblock Phys. Rev. B {\bf 84} 085146 (2011).

\bibitem{santos11}
L.~F.~Santos and A.~Mitra
\newblock Phys. Rev. E {\bf 84} 016206 (2011).

\bibitem{jesenko11}
S.~Jesenko and M.~$\check{\mbox{Z}}$nidari$\check{\mbox{c}}$
\newblock arxiv.1105.6340v1 (unpublished).

\bibitem{kollath05a}
C.~Kollath, U.~Schollw\"ock, and W.~Zwerger,
\newblock Phys. Rev. Lett. {\bf 95}, 176401 (2005).

\bibitem{polini07}
M.~Polini and G.~Vignale,
\newblock Phys. Rev. Lett. {\bf 98}, 266403 (2007).

\bibitem{jowi}
P.~Jordan and E.~Wigner
\newblock Z. Phys. {\bf 47} 631 (1928)

\bibitem{hess07}
C.~Hess, H.~ElHaes, A.~Waske, B.~B\"uchner, C.~Sekar, G.~Krabbes,
  F.~Heidrich-Meisner, and W.~Brenig,
\newblock Phys.\ Rev.\ Lett. {\bf 98}, 027201 (2007).

\bibitem{sologubenko07a}
A.~V. Sologubenko, T.~Lorenz, H.~R. Ott, and A.~Freimuth,
\newblock J. Low Temp. Phys. {\bf 147}, 387 (2007).

\bibitem{hess01}
C.~Hess, C.~Baumann, U.~Ammerahl, B.~B\"uchner, F.~Heidrich-Meisner, W.~Brenig,
  and A.~Revcolevschi,
\newblock Phys.\ Rev.\ B {\bf 64}, 184305 (2001).

\bibitem{sologubenko00}
A.~V. Sologubenko, K.~Gianno, H.~R. Ott, U.~Ammerahl, and A.~Revcolevschi,
\newblock Phys.\ Rev.\ Lett. {\bf 84}, 2714 (2000).

\bibitem{hlubek10}
N.~Hlubek, P.~Ribeiro, R.~Saint-Martin, A.~Revcolevschi, G.~Roth, G.~Behr,
  B.~B\"uchner, and C.~Hess,
\newblock Phys. Rev. B {\bf 81}, 020405 (2010).

\bibitem{otter09}
M.~Otter, V.~Krasnikov, D.~Fishman, M.~Pshenichnikov, R.~Saint-Martin,
  A.~Revcolevschi, and P.~van Loodsrecht,
\newblock J. Mag. Mag. Mat. {\bf 321}, 796 (2009).

\bibitem{daley04}
A.~Daley, C.~Kollath, U.~Schollw\"ock, and G.~Vidal,
\newblock J. Stat. Mech.: Theory Exp. (2004) P04005 .

\bibitem{white04}
S.~R. White and A.~E. Feiguin,
\newblock Phys.\ Rev.\ Lett. {\bf 93}, 076401 (2004).

\bibitem{vidal04}
G.~Vidal,
\newblock Phys. Rev. Lett. {\bf 93}, 040502 (2004).

\bibitem{schollwoeck05}
U.~Schollw\"ock,
\newblock Rev. Mod. Phys. {\bf 77}, 259 (2005).

\bibitem{schollwoeck11}
U.~Schollw\"ock,
\newblock Ann. Phys. (NY) {\bf 326}, 96 (2011)
 
\bibitem{zotos97}
X.~Zotos, F.~Naef, and P.~Prelov{\v{s}}ek,
\newblock Phys.\ Rev.\ B {\bf 55}, 11029 (1997).

\bibitem{kluemper02}
A.~Kl\"umper and K.~Sakai,
\newblock J. Phys. A {\bf 35}, 2173 (2002).

\bibitem{sakai03}
K.~Sakai and A.~Kl\"umper,
\newblock J. Phys. A {\bf 36}, 11617 (2003).

\bibitem{hm02}
F.~Heidrich-Meisner, A.~Honecker, D.~C. Cabra, and W.~Brenig,
\newblock Phys.\ Rev.\ B {\bf 66}, 140406(R) (2002).

\bibitem{hm03}
F.~Heidrich-Meisner, A.~Honecker, D.~C. Cabra, and W.~Brenig,
\newblock Phys. Rev. B {\bf 68}, 134436 (2003).

\bibitem{narozhny98}
B.~N. Narozhny, A.~J. Millis, and N.~Andrei,
\newblock Phys. Rev. B {\bf 58}, R2921 (1998).

\bibitem{zotos99}
X.~Zotos,
\newblock Phys.\ Rev.\ Lett. {\bf 82}, 1764 (1999).

\bibitem{zotos04}
X.~Zotos,
\newblock Phys. Rev. Lett. {\bf 92}, 067202 (2004).

\bibitem{prelovsek04}
P.~Prelov\ifmmode \check{s}\else\v{s}\fi{}ek, S.~El~Shawish, X.~Zotos and M.~Long,
\newblock Phys. Rev. B {\bf 70}, 205129 (2004)

\bibitem{benz05}
J.~Benz, T.~Fukui, A.~Kl\"umper, and C.~Scheeren,
\newblock J. Phys. Soc. Jpn. Suppl. {\bf 74}, 181 (2005).

\bibitem{jung06}
P.~Jung, R.~W. Helmes, and A.~Rosch,
\newblock Phys.\ Rev.\ Lett. {\bf 96}, 067202 (2006).

\bibitem{sirker09}
J.~Sirker, R.~G. Pereira, and I.~Affleck,
\newblock Phys. Rev. Lett. {\bf 103}, 216602 (2009).

\bibitem{steinigeweg09}
R.~Steinigeweg and J.~Gemmer
\newblock Phys. Rev. B, {\bf 80}, 184402 (2009)

\bibitem{grossjohann10}
S.~Grossjohann and W.~Brenig,
\newblock Phys. Rev. B {\bf 81}, 012404 (2010).

\bibitem{steinigeweg11}
R.~Steinigeweg
\newblock Phys. Rev. E {\bf 84 } 011136 (2011).
\bibitem{steinigeweg11a}
R.~Steinigeweg and W.~Brenig,
\newblock arXiv.1107.3103 (unpublished).

\bibitem{herbrych11}
J.~Herbrych, P.~Prelov\ifmmode \check{s}\else\v{s}\fi{}ek, X.~Zotos,
\newblock arXiv.1107.3027 (unpublished).

\bibitem{prosen11}
T.~Prosen,
\newblock Phys. Rev. Lett. {\bf 106}, 217206 (2011).

\bibitem{znidaric11}
M.~\ifmmode \check{Z}\else
  \v{Z}\fi{}nidari\ifmmode~\check{c}\else \v{c}\fi{},
\newblock Phys. Rev. Lett. {\bf 106}, 220601 (2011).

\bibitem{sirker11}
J.~Sirker, R.~G. Pereira, and I.~Affleck,
\newblock Phys. Rev. B {\bf 83}, 035115 (2011).

\bibitem{caux11}
J.-S. Caux and J.~Mossel,
\newblock J. Stat. Mech. ({\bf2011}) P02023.

\bibitem{prosen09}
T.~Prosen and M.~\ifmmode \check{Z}\else
  \v{Z}\fi{}nidari\ifmmode~\check{c}\else \v{c}\fi{},
\newblock J. Stat. Mech: Theor. Exp. ({\bf 2009}) P02035 .

\bibitem{benenti09}
G.~Benenti, G.~Casati, T.~Prosen, and D.~Rossini,
\newblock EPL {\bf 85}, 37001 (2009).

\bibitem{benenti09a}
G.~Benenti, G.~Casati, T.~Prosen, D.~Rossini, and M.~\ifmmode \check{Z}\else
  \v{Z}\fi{}nidari\ifmmode~\check{c}\else \v{c}\fi{},
\newblock Phys. Rev. B {\bf 80}, 035110 (2009).

\bibitem{calabrese07}
P.~Calabrese and J.~Cardy,
\newblock J. Stat. Mech. ({\bf 2007}) P10004.


\bibitem{rigol08}
M.~Rigol and B.~S.~Shastry,
\newblock Phys. Rev. B {\bf 77}, 161101(R) (2008)

\bibitem{chandrasekhar43}
S.~Chandrasekhar
\newblock Rev. Mod. Phys. {\bf 15}, 1 (1943).

\bibitem{steinigeweg10}
R.~Steinigeweg and R.~Schnalle
\newblock Phys. Rev. E {\bf 82}, 040103(R) (2010)

\bibitem{solyom79}
J.~Solyom, Adv. Phys. {\bf 28}, 209 (1979).

\bibitem{vondelft98}
J.~von Delft and H.~Schoeller,
\newblock Ann.~Phys.~(Leipzig) {\bf 7}, 225 (1998).

\bibitem{cloizeaux66}
J.~des Cloizeaux and M.~Gaudin,
\newblock J. Math. Phys. {\bf 7}, 1384 (1966).

\bibitem{hulthen38}
L.~Hulth\ifmmode \check{e}\else\v{e}\fi{}n,
\newblock Arkiv. Mat. Astron. Fysik 26A No. 11 (1938)

\bibitem{park86}
T.~Park and J.~Light,
\newblock J. Chem. Phys {\bf 85}, 5870 (1986).

\bibitem{hochbruck97}
M.~Hochbruck and C.~Lubich,
\newblock SIAM J. Numer. Anal. {\bf 34}, 1911 (1997).

\bibitem{mcculloch02}
I.~McCulloch and M.~Gulasci,
\newblock EPL {\bf 57}, 852 (2002).

\end{thebibliography}
\end{document}